\documentclass[prl,aps,twocolumn,amsfonts,showpacs,superscriptaddress,preprintnumbers,footinbib,longbibliography]{revtex4-1}

\usepackage{graphicx}
\usepackage{xcolor}
\usepackage{rotating}
\usepackage{bm,amsmath,amssymb}
\usepackage[utf8]{inputenc}
\usepackage{footmisc}

\usepackage[colorlinks=true, urlcolor=blue, linkcolor=blue, citecolor=blue, hyperindex=true, linktocpage=true]{hyperref}

\newcommand{\nn}{\nonumber\\}

\def\CC{\mathcal{C}}
\def\CD{\mathcal{D}}

\def\CL{\mathcal{L}}

\def\CP{\mathcal{P}}

\def\CT{\mathcal{T}}

\def\sig{{\sigma}}
\def\Lam{{\Lambda}}
\def\Ga{{\Gamma}}
\def\le{\left}
\def\ri{\right}

\newcommand{\pt}{\partial}
\newcommand\para{{\parallel}}
\newcommand\ov{\over}

\newcommand{\vx}{\mathbf{x}}
\newcommand{\vv}{\mathbf{v}}
\newcommand{\vk}{\mathbf{k}}

\newcommand\Th{{\Theta}}
\newcommand\nab{{\nabla}}

 \newcommand{\bB}{\mathbf{B}}
  \newcommand{\bE}{\mathbf{E}}

\def\be{\begin{equation}}
\def\ee{\end{equation}}

\def\bea{\begin{eqnarray}}
\def\eea{\end{eqnarray}}
\newcommand{\bma}{\le(\begin{matrix}}
\newcommand{\ema}{\end{matrix}\ri)}

\newcommand{\bega}{\begin{gather}}

\newcommand\ha{{1 \ov 2}}
\newcommand\p{\ensuremath{\partial}}
\newcommand{\ep}{\epsilon}
\newcommand{\de}{\delta}
\newcommand{\om}{\omega}

\newcommand{\bk}{\mathbf{k}}

\newcommand{\mnras}{Monthly Notices of the Royal Astronomical Society}
\newcommand{\aap}{Astronomy and Astrophysics}

\begin{document}
\preprint{MIT-CTP/5442}

\title{Strong-field magnetohydrodynamics for neutron stars}

\author{Shreya Vardhan}
\email{vardhan@stanford.edu}
\affiliation{Stanford Institute for Theoretical Physics, Stanford University, Stanford, CA 94305 
}

\author{Sa\v{s}o Grozdanov}
\email{saso.grozdanov@ed.ac.uk}
\affiliation{Higgs Centre for Theoretical Physics, University of Edinburgh, Edinburgh, EH8 9YL, Scotland, 
}

\affiliation{Faculty of Mathematics and Physics, University of Ljubljana, Jadranska ulica 19, SI-1000 Ljubljana, Slovenia
}

\author{Samuel Leutheusser}
\email{sl9535@princeton.edu}
\affiliation{Princeton Gravity Initiative, Princeton University, Princeton NJ 08544, USA 
}

\author{Hong Liu}
\email{hong\_liu@mit.edu}
\affiliation{Center for Theoretical Physics, MIT, Cambridge, MA 02139, USA 
}

\begin{abstract}
We present a formulation of magnetohydrodynamics which can be used to describe the evolution of strong magnetic fields in neutron star interiors. Our approach is based on viewing magnetohydrodynamics as a theory with a one-form global symmetry and developing an effective field theory for the hydrodynamic modes associated with this symmetry. In the regime where the local velocity and temperature variations can be neglected, we derive the most general constitutive relation consistent with symmetry constraints for the electric field in the presence of a strong magnetic field. This constitutive relation not only reproduces the phenomena of Ohmic decay, ambipolar diffusion, and Hall drift derived in a phenomenological model by  Goldreich and Reisenegger, but also reveals new terms in the evolution of the magnetic field which cannot easily be seen from such microscopic models.    
This formulation gives predictions for novel diffusion behaviors of small perturbations around  
a constant background magnetic field, and for the two-point correlation functions among various components of the electric and magnetic fields.  
\end{abstract}
\maketitle

{\bf Introduction.}---
Hydrodynamics provides a universal description of 
an interacting many-body system in the large distance and long time limit
by capturing the effective dynamics of conserved quantities. 
For an electrically conducting fluid with a magnetic field, it is necessary to incorporate electromagnetism in describing the 
motion of the fluid, leading to magnetohydrodynamics (MHD). 
MHD plays an important role in many disciplines including plasma physics, astrophysics, and cosmology. 
In this paper, we provide a new formulation of MHD at strong magnetic fields for neutron stars, which 
 exhibit some of the strongest (up to $10^{11}$ Tesla) and most dramatically variable magnetic fields in the universe (see Refs.~\cite{Lattimer:2004pg,Kaspi:2017fwg,harding2006physics}). We also discuss some immediate implications of the new theory for magnetic diffusion. 

The dynamical variables of MHD for a neutral fluid include the local velocity $\vv (x)$, local temperature $ T(x)$, and magnetic field
$\bB (x)$. Here bold-face letters denote spatial vectors.  The equations of motion consist of the conservation of the energy-momentum tensor $\p_\mu T^{\mu \nu} =0$ \footnote{We use $\mu, \nu, \cdots$ to denote spacetime indices and $i,j, \cdots$ to denote spatial indices.}, and two of the Maxwell equations,
\be\label{mhdeq}
\nabla \cdot {\bf B} = 0, \quad
\partial_t {\bf B} = - \nabla \times {\bf E} .
\ee
The components of $T^{\mu \nu}$ and $ {\bf E} $ are expressed in terms of the hydrodynamical variables via constitutive relations. 
In the standard formulation of MHD, such relations are written on phenomenological grounds and become ambiguous in the presence of a strong magnetic field.

In applications to a neutron star, a frequently-used approximation to illustrate some of the main physical effects is to take $\vv (x) =0$ \footnote{This amounts to assuming that the neutron background is static. This assumption has been relaxed in recent works (see e.g.~\cite{ofengeim, castillovelocity}), leading to faster evolution of the magnetic field, but it can still be used as a first approximation to illustrate the main physical effects.}  and $T (x) = {\rm const}$.
In~\cite{1992ApJ...395..250G}, a phenomenological model for neutron stars was used to derive an expression for {\bf E} in terms of {\bf B} as well as fast-evolving microscopic variables such as the densities of electrons and protons. 
The results of~\cite{1992ApJ...395..250G}  have been used to formulate MHD equations in a variety of contexts, both in neutron star physics \cite{crust, core, combined, Lattimer:2004pg,Kaspi:2017fwg,harding2006physics,thompson1993neutron,thompson1995soft,urpin1999magnetohydrodynamic,Cumming:2004mf,cho2004anisotropy,Reisenegger:2005nb,Mereghetti:2015asa,Passamonti:2016nmf,Pons_2019} and in astrophysics more generally \cite{Brandenburg1994TheFO,brandenburg2005astrophysical,balbus2009magnetohydrodynamics}. Despite these successes, the inadequate knowledge of particle interactions inside neutron stars makes it difficult to assess the accuracy of such formulations or to improve upon them systematically using phenomenological approaches. 

In this paper, we provide a new formulation of strong-field MHD based on effective field theory, which makes it possible to derive the most general constitutive relations consistent with the symmetries of the system, regardless of the details of microscopic interactions~(such as whether weak interactions are included or not). For illustration of the formalism,  we will use the static neutron approximation, and neglect possible neutron superfluidity and proton superconductivity~\cite{10.1111/j.1365-2966.2010.17484.x, dommes}. Our theory is valid in the regime where all other physical quantities relax much faster than $\bB$. Our main result is a constitutive relation for ${\bf E}$ in terms of ${\bf B}$ given in \eqref{ouC}, which together with \eqref{mhdeq} gives a closed equation for the evolution of ${\bf B}$. In this paper, we outline the derivation of this result and its consequences. A detailed exposition of the effective field theory is provided in~\cite{long}.

Our formulation combines two significant recent  developments. 
The first ingredient is a new formulation of hydrodynamics  as the low energy effective field theory (EFT) of a general many-body system~\cite{Crossley:2015evo,Glorioso:2017fpd} (see~\cite{Glorioso:2018wxw} for a review and also~\cite{Grozdanov:2013dba,Haehl:2015uoc,Jensen:2017kzi}). In addition to providing a systematic framework for treating hydrodynamical  fluctuations, the new EFT approach also has a number of advantages at the level of equations of motion compared with the traditional approach. It provides a systematic derivation of the constitutive relations without the need of imposing any phenomenological constraints by hand, and introduces an alternative set of dynamical variables that are often more convenient than traditional ones. The second ingredient is the important observation of~\cite{Grozdanov:2016tdf} that MHD can be formulated as an effective field theory for a system with a higher-form global symmetry~\cite{Gaiotto:2014kfa}. See \cite{Hernandez:2017mch,Grozdanov:2017kyl,Glorioso:2018kcp,Armas:2018zbe,Gralla:2018kif,Benenowski:2019ule,Landry:2021kko,Armas:2022wvb} for various developments. This makes it possible to formulate MHD based on symmetries alone.

{\bf MHD for neutron stars.}---The basic idea behind the formulation of MHD in~\cite{Grozdanov:2016tdf} is that the MHD equations~\eqref{mhdeq} can be written as the conservation equations for a two-form current 
$J^{\mu \nu} = - J^{\nu \mu}$:
\begin{align}\label{J_conservation}
\partial_\mu J^{\mu\nu} = 0, \quad  J^{0i} = B^i , \quad J^{ij} =- \epsilon^{ijk} E^k , 
\end{align}
i.e., $J^{\mu\nu} =\tilde F^{\mu\nu} = \frac{1}{2} \epsilon^{\mu\nu\rho\sigma} F_{\rho\sigma}$. Accordingly, equations~\eqref{mhdeq} can be interpreted as resulting from an underlying one-form symmetry. This realization is not simply a renaming of equations~\eqref{mhdeq}, but provides a powerful symmetry principle for formulating MHD. In particular, the choice of dynamical variables and the structure of the constitutive relation are dictated by this symmetry principle. 

An effective action for a dissipative hydrodynamic system can be formulated in terms of the closed-time-path (CTP) or Schwinger-Keldysh formalism (for a review of these methods, see~\cite{Glorioso:2018wxw}). A key ingredient of the formulation of~\cite{Crossley:2015evo,Glorioso:2017fpd} is the fact that the dynamical hydrodynamic variables are the Stueckelberg fields associated with the global symmetries responsible for the conservation laws. For the one-form symmetry relevant for MHD, this is a vector field $A_\mu$, which is a collective effective field describing the long-distance and long-time behavior (not the microscopic electromagnetic potential for $\bf E$ and $\bf B$ fields of the system). 

The hydrodynamic effective action $S_{\rm EFT}$ for MHD is then the most general action written in terms of the combination $G_{\mu\nu} \equiv b_{\mu\nu} + \partial_\mu A_{\nu} - \partial_\nu A_{\mu}$ that is invariant under various symmetries. Here, for later convenience, we have also turned on an external source $b_{\mu \nu}$  for the two-form current $J^{\mu \nu}$. The symmetries to be imposed can be separated into those universal for all hydrodynamic systems~\cite{Crossley:2015evo,Glorioso:2017fpd} and those specific to the system under study. A neutron star can be regarded as a neutral plasma consisting of electrons, protons, and neutrons. In this system, the energy is high enough that electrons can be treated as relativistic, but not high enough for the existence of positrons. Thus, the charge conjugation $\CC$ is badly broken. Weak interactions also break parity $\CP$. See Section I of the Supplementary Material (SM) for the explicit form of the action. Here, we give the final form of the constitutive relations for $J^{\mu \nu}$ at the lowest dissipative order:
\begin{align}\label{j10}
J^{0i} =& \, a \, G_{0i}  
+ m \, \delta_{ij} {\epsilon}_{kln} G_{0n} H_{jkl} ,
 \\
 \label{JC-4-20} 
 J^{ij} = &
        -2 \beta_0 (d \, \delta_{ik} \delta_{jl}  + \tilde d \, \epsilon_{ijm} {\epsilon}_{kln} G_{0m} G_{0n}) \pt_0 G_{kl} \cr
 &-  m \,  \epsilon_{lij}   \pt_l G_{0k}^2  + 2 p\,  \delta_{k[i} \epsilon_{j]ln} 
     G_{0n} \partial_0 G_{kl}  ,
\end{align}
where $H_{\mu \nu \lambda}  = \partial_{\mu} b_{ \nu \lambda} + \partial_{\lambda} b_{\mu \nu} + \partial_{\nu} b_{\lambda\mu}$ and $[ij] \equiv {1 \over 2}  (ij - ji)$. All coefficients $a, d, \tilde d, m, p$ should be understood as functions of $G_{0i}^2$,  and they satisfy the constraints
\be 
a \geq 0, \quad  d \geq 0 , \quad 
d + 2 \tilde d \,  G_{r0i}^2 \geq 0  . \label{ineq}
\ee

We can write~\eqref{j10}--\eqref{JC-4-20} in a more conventional form by solving for $G_{0i}$ in terms of $J^{0i} = B^i$. Then on setting $b_{\mu \nu}=0$~\footnote{Note that $\partial_0 G_{ij}= H_{0ij}+ \partial_i G_{0j}-\partial_j G_{0i}$.},~\eqref{JC-4-20} can be written as the constitutive relation for the electric field:
\begin{align}\label{ouC}
 {\bf E} &= \eta \, {\bf j} + c_a \, ({\bf j} \times {\bf B})\times{ \bf B} + c_H \, {\bf j} \times {\bf B} \\
 &- c_H (\nabla \ln a({\bf B}^2) \times {\bf B}) \times {\bf B} - c_{\eta} \nabla  \ln a({\bf B}^2) \times {\bf B}  +  \nabla f ,  \nonumber
 \end{align} 
where ${\bf j} \equiv \nabla \times {\bf B}$, and we have introduced a function $f$ of ${\bf B}^2$ such that $m=f'$. The coefficients are 
\begin{equation} 
\label{idf}
\eta =  c_\eta+ c_a \bB^2,~c_\eta = \frac{2 \beta_0 d}{a} , ~c_a =  \frac{4 \beta_0 \tilde d}{a^3},  ~ c_H = - \frac{p}{a^2} ,
\end{equation} 
where $\beta_0 = 1 / T$ is the constant inverse temperature. Now all coefficients should be understood as functions of $\bB^2$. We will see below that $a (\bB^2)$ can be interpreted as a one-form susceptibility in the linear response theory around a constant background magnetic field ${\bf B}$, and must therefore be non-negative. 
Note that $\nabla \ln a({\bf B}^2) = \frac{a'}{a}\nabla {\bf B}^2$, and, hence, is proportional to the gradient of the magnetic energy density. Coefficients $c_\eta$ and $\eta$ will be interpreted as magnetic ``diffusivities.''
From~\eqref{ineq},  we must have
\be \label{ineq1} 
\eta \geq 0, \quad c_\eta \geq 0  .
\ee
Note that \eqref{ouC} can also be written in terms of the three independent coefficients $c_{\eta}$, $c_a$ and $c_H$ as 
\begin{gather}
{\bf E} = c_{\eta}\,  {\bf j}_B +c_a \, ({\bf B} \cdot {\bf j}) \, {\bf B} + c_H \, {\bf j}_B\times {\bf B} + \nabla f  , \\
{\bf j}_{B} = {\bf j} - \nabla \ln a({\bf B}^2) \times {\bf B}  . 
\end{gather}

The first line of \eqref{ouC} includes each of the explicitly ${\bf B}$-dependent terms derived in equation (13) of \cite{1992ApJ...395..250G}, that lead to behaviors in the magnetic field evolution known as Ohmic decay, ambipolar diffusion, and Hall drift respectively. We would, however, like to emphasize two significant ways in which this equation provides a generalization of the formalism of \cite{1992ApJ...395..250G}. Firstly, the terms in the second line of \eqref{ouC} are not explicitly present in \cite{1992ApJ...395..250G}. The physical origin of these terms has been clarified in~\cite{Landry:2022nog} and their implications will be further elaborated in~\cite{LandryLiu}. More explicitly, $a({\bf B}^2)$ can be identified as the magnetic permeability  of the system. Thus the second line of~\eqref{ouC} captures implications of ${\bf H} = {{\bf B} \ov a ({\bf B}^2)}$ being different from $\bB$~\footnote{In this equation, ${\bf H}$ refers to the external magnetic field, which does not appear explicitly in our formalism, and in particular should not be confused with the higher form field $H_{ijk}$.}. Note that these terms only depend on the gradient of the magnetic permeability.

Another way in which the above constitutive relation generalizes the one from \cite{1992ApJ...395..250G} is that  the coefficients $\eta$, $c_a$, and $c_H$ are all functions of ${\bf B}$ in \eqref{ouC}. Note also that the value of $c_a$, which is negative in~\cite{1992ApJ...395..250G}, can in principle have either sign here. If negative, its absolute value is bounded by $|c_a| \bB^2 < c_\eta$ due to \eqref{ineq1}.  

Here  we see the major advantages of our formulation. The constitutive relation of $\bE$ in terms of $\bB$ is rather complicated, and there appears to be no general principle to write it down directly. But in terms of the new dynamical variables $A_\mu$ (or equivalently $G_{0i}$), the fields $\bE$ and $\bB$ are given in a parametric form in~\eqref{j10}--\eqref{JC-4-20}, which can be written down systematically and straightforwardly based on symmetries and derivative expansions. We believe the parametric form~\eqref{j10}--\eqref{JC-4-20} is likely also more convenient for nonlinear numerical simulations.

{\bf Magnetic diffusion with a strong background magnetic field.}---To develop intuition for the physics encoded in~\eqref{j10}--\eqref{JC-4-20} (or equivalently,~\eqref{ouC}), we consider the behavior of small perturbations around a constant $\bf B$ along the $z$ direction with magnitude $B_0$. Such a constant magnetic field configuration can be conveniently thought of as being due to a constant background `chemical potential' $b_{0z} = \mu$ with $B_0 = a (\mu^2) \mu$. For small perturbations around such a configuration, we can write
\begin{equation}\label{Br0i-Def}
\!\! G_{\mu \nu} =2 \mu \delta_{0 [\mu} \delta_{\nu] z}     + f_{\mu \nu} , \; \, f_{\mu 
\nu} =  b_{\mu \nu} +  \partial_\mu A_{\nu} - \partial_\nu A_{\mu} ,
\end{equation}
where we have also introduced an infinitesimal external source $b_{\mu \nu}$ and we will work to linear order in both $b_{\mu \nu}$ and $f_{\mu \nu}$. Expanding~\eqref{j10}--\eqref{JC-4-20} in small $f_{\mu \nu}$ we then find
\begin{align}\label{J_exp}
J^{0z} &= B_0 +  \chi_\parallel f_{0z} +2 m \mu H_{xyz} , \quad J^{0a} = \chi_\perp  f_{0a}, \\
J^{za} &=- ( \sigma_1^\perp  \delta_{ab} + \sigma_2^\perp \epsilon_{ab}) (H_{0zb} + \pt_z f_{0b}) \cr
&+ \left[\sigma_1^\perp  \delta_{ab} + (\sigma_2^\perp- 2 m \mu  ) \epsilon_{ab}  \right] \pt_b f_{0z}, \\
J^{xy} &= -\sigma^\para (H_{0xy} + K_{xy}) - 2 m \mu \p_z f_{0z} ,
\label{hek}
\end{align}
where subscripts $a,b$ index $x, y$ directions, and
\begin{align}\label{p1}
& \chi_\para = a + \tilde a \mu^2, \quad \chi_\perp = a , \quad  \tilde{a}= 2 a' (\mu^2) , \\
\label{p2} 
&\sigma_1^\perp = c_\eta \chi_\perp
\quad \sigma_2^\perp ={B_0 } c_H  \chi_\perp, \quad \sigma^\para = \eta \chi_\perp, \\
& \qquad K_{xy} \equiv \pt_x f_{0y} - \pt_y f_{0x} ,
\label{p3}
\end{align}
where all coefficients should be evaluated at $ {\bf B}^2 = B_0^2$ through $\mu$. As discussed in Section I of SM, systems with charge conjugation symmetry have $p=0$, and hence $\sig_2^\perp =0$. We stress that $\tilde a$ arises from $\bB$-dependence of the coefficient $a$, and hence cannot be seen naively from the formulation of~\cite{1992ApJ...395..250G}, where all coefficients appear to be $\bB$-independent.

\begin{figure}[ht!] 
\centering
\includegraphics[width=0.4\textwidth]{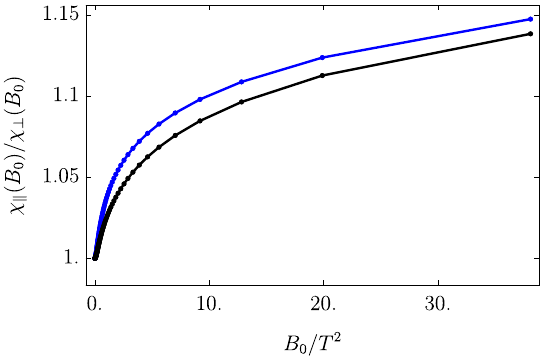}
\caption{Ratio of one-form susceptibilities $\chi_\para/\chi_\perp$ computed numerically from a holographic MHD model \cite{Grozdanov:2017kyl} as a function of equilibrium $B_0/T^2$. Black and blue dots correspond smaller and larger choices of the electromagnetic coupling, respectively.}
\label{fig:Hol}
\end{figure}

The functions $\chi_\perp$ and $\chi_\para$ can be interpreted as ``one-form susceptibilities'', defined by the Kubo formulas
\be \label{y2t}
\chi_\para = \lim_{\vk \to 0} \lim_{\om \to 0}  G^R(B_z, B_z), \quad \chi_\perp = \lim_{\vk \to 0} \lim_{\om \to 0}  G^R(B_x, B_x), 
\ee
where $G^R(B_z, B_z)$ denotes the retarded correlation functions of $B_z$ and $B_z$, and similarly for other components~(see Section II of SM for details). From thermodynamic stability we should have $\chi_\perp \geq 0, \chi_\para \geq 0$.  {In the linear response theory derived from~\cite{1992ApJ...395..250G}, we have $\chi_{\para}=\chi_{\perp}$ since $\tilde a=0$, but more generally, these susceptibilities should be different.} In Figure~\ref{fig:Hol}, we show their ratio in a holographic MHD model \cite{Grozdanov:2017kyl} (see also \cite{Hofman:2017vwr}).

The $\sig$'s can be interpreted as  magnetic ``viscosities'', defined by the Kubo formulas 
\bega \label{y1t}
\sig^\para = \lim_{\om \to 0} \lim_{\vk \to 0} {1 \ov i \om} G^R(E_z, E_z) , \\ 
\sig^\perp_1 = \lim_{\om \to 0} \lim_{\vk \to 0} {1 \ov i \om} G^R(E_x, E_x) , \\
\sig^\perp_2 = \lim_{\om \to 0} \lim_{\vk \to 0} {1 \ov i \om} G^R(E_x, E_y) ,
\label{y7t}
\end{gather} 
where $\sig^\perp_2$ can be understood as a ``Hall'' magnetic viscosity. We stress the differences in the order of limits between~\eqref{y2t} and~\eqref{y1t}--\eqref{y7t}. Comparing~\eqref{ineq1} and~\eqref{p2},
\be 
\sigma^{\para} \geq 0, \quad \sigma_1^{\perp} \geq 0, 
\ee
but $\sigma_2^{\perp}$ can have either sign. Note that $\sig^\para  -\sig^\perp_1 \propto c_a$, and thus the effect of the ambipolar term at this order is to generate anisotropy in the magnetic viscosities.

The equations of motion following from~\eqref{J_exp}--\eqref{hek} are a set of coupled diffusion equations, which give the following dispersion relations (see Section II of SM for details): 
\begin{align}\label{MHD_FullDispRel}
&\omega = - i  {\sig^{\perp}_1 \ov \chi_\perp} k_z^2 -  \frac{i}{2} \le({\sig^\para \ov \chi_\perp} + {\sig^{\perp}_1 \ov \chi_\para}   \ri) k_\perp^2 \\
&\mp \frac{i}{2} \sqrt{    \le({\sig^\para \ov \chi_\perp} - {\sig^{\perp}_1 \ov \chi_\para}   \ri)^2 k_\perp^4   - 4  \left(   \frac{ \sig^\perp_2}{  \chi_\perp} \right)^2 \le(k_z^2 + \frac{\chi_\perp}{\chi_\para} k_\perp^2  \ri)  
 k_z^2  }  , \nonumber
\end{align}
where $k_\perp^2 = k_x^2 + k_y^2$. 
The special case of \eqref{MHD_FullDispRel} with $\chi_{\para}=\chi_{\perp}$ was derived from the formulation of \cite{1992ApJ...395..250G} in \cite{vigano2}, but the general case is a new prediction from our formalism.
Note also the special cases  
\begin{align}
k_\perp =0: & \quad \om = - {i \ov \chi_\perp} (\sig_1^\perp \pm i \sig_2^\perp) k_z^2,  \label{yeun1} \\
k_z =0:&\quad  \om = - i {\sig^\para \ov \chi_\perp} k_\perp^2 \;\; \text{or} \;\;  \om = - i  {\sig^{\perp}_1 \ov \chi_\para} k_\perp^2  . \label{yeun2}
\end{align} 

\begin{figure}[t] 
\centering
\includegraphics[width=0.238\textwidth]{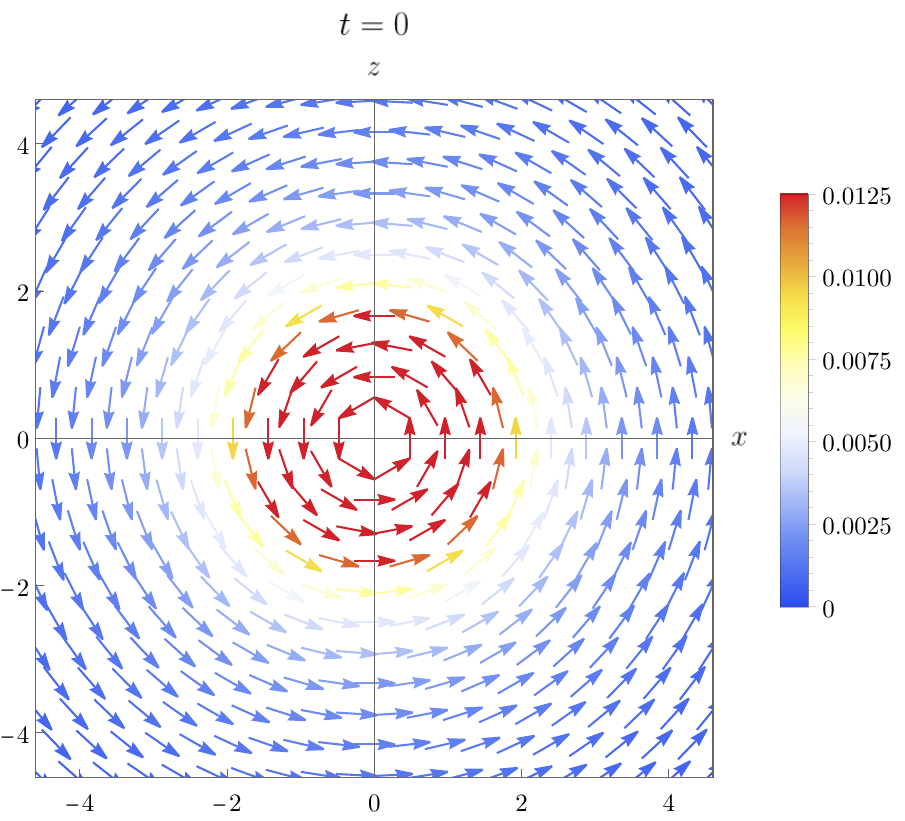}\\
\includegraphics[width=0.238\textwidth]{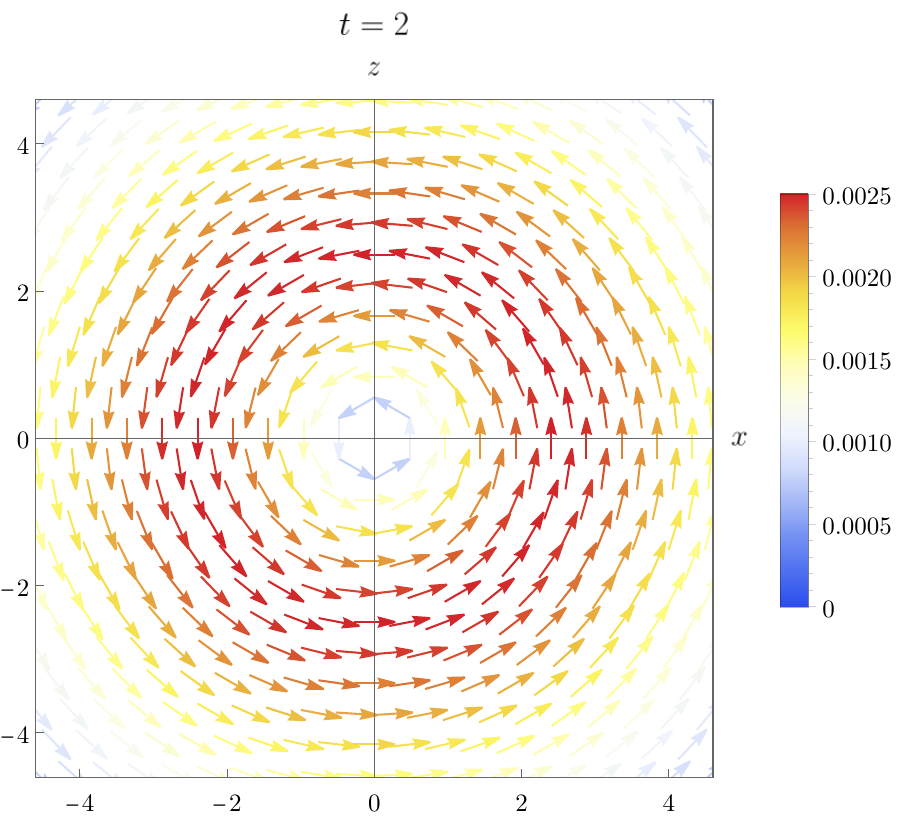}
\includegraphics[width=0.238\textwidth]{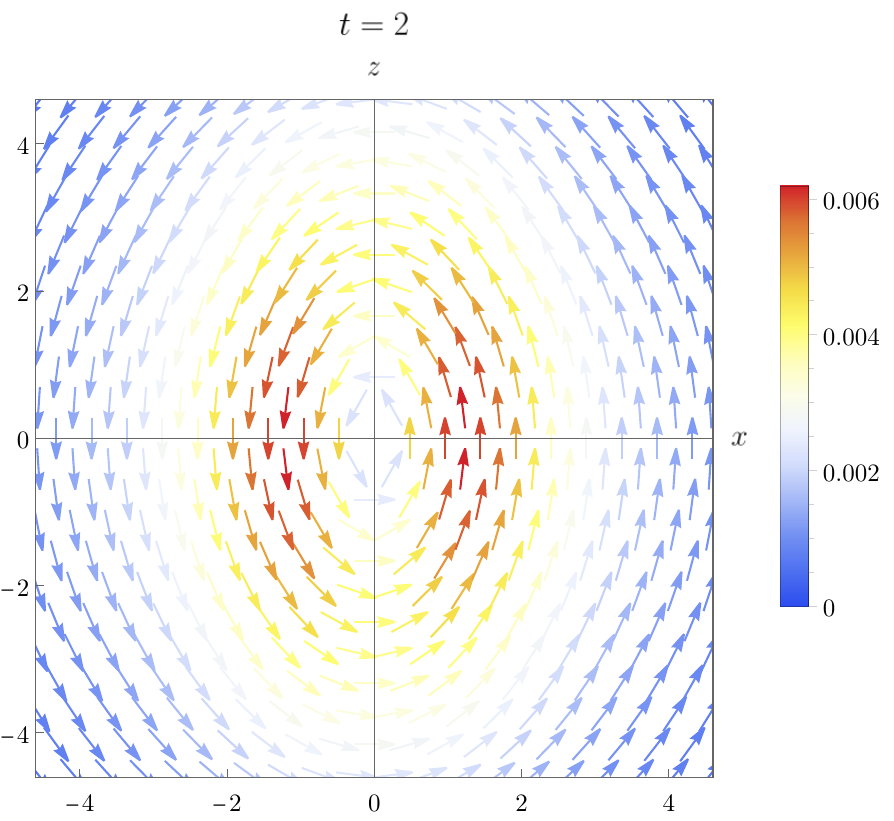}
\caption{Evolution of the initial magnetic field configuration 
${\bf B}(0, \vx)
= \frac{\sqrt{\pi} r}{(2\pi)^2} e^{-\frac{r^2}{2}} \left[ I_0\left( \frac{r^2}{2}\right) - I_1\left( \frac{r^2}{2} \right) \right] { \hat \phi}$, 
where $r, \phi$ are the polar coordinates in the $x-z$ plane, and $I_n(r)$ is the modified Bessel function of the first kind. We take $\sigma_1^{\perp}=1$. The top figure shows the initial condition. The bottom left figure shows the configuration at $t=2$ with $\chi_{\parallel}=\chi_{\perp}=1$, while the right figure shows the configuration at $t=2$ with $\chi_{\parallel}= 1$ and $\chi_{\perp}=8$. The color bars indicate the magnitude of ${\bf B}$.}
\label{fig:iso}
\end{figure}

The above dispersion relations lead to interesting patterns in the diffusion of magnetic field lines.
First consider $\sig_2^\perp =0$ (note that this approximation is valid in certain regions of the neutron star such as the core \cite{core}), in which case~\eqref{MHD_FullDispRel} becomes 
\be
\om =-{i \ov \chi_\perp}  \le(\sig^{\perp}_1 k_z^2+ \sig^{\para}  k_\perp^2 \ri) \;\; \text{or} \;\; 
\om =-i \sig^{\perp}_1 \le({ k_z^2 \ov \chi_\perp}   +  {k_\perp^2 \ov \chi_\para}   \ri)  .
\ee
Both modes diffuse anisotropically between $z$ and some direction in the $x-y$ plane. 
When either $\sig^{\perp}_1 = \sig^{\para} $ (i.e.,~$c_a =0$) or $ \chi_\perp = \chi_\para$, one of the modes becomes isotropic. 
See Fig.~\ref{fig:iso} for an example of the qualitatively different evolutions with $\chi_{\perp}=\chi_{\para}$, corresponding to the formalism of \cite{1992ApJ...395..250G},  and the generic case of $\chi_{\perp}\neq \chi_{\para}$~\footnote{Note that we need an initial configuration which involves both $k_z$ and $k_\perp$
to probe this difference}.

\begin{figure}[t]
\centering
\includegraphics[width=0.238\textwidth]{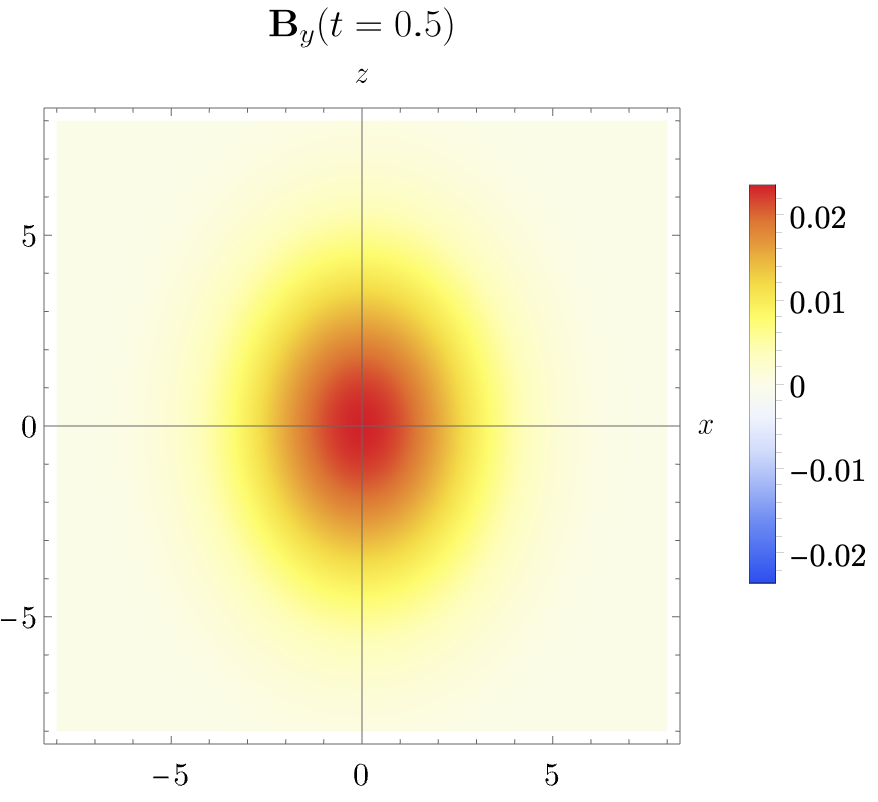}\\
\includegraphics[width=0.238\textwidth]{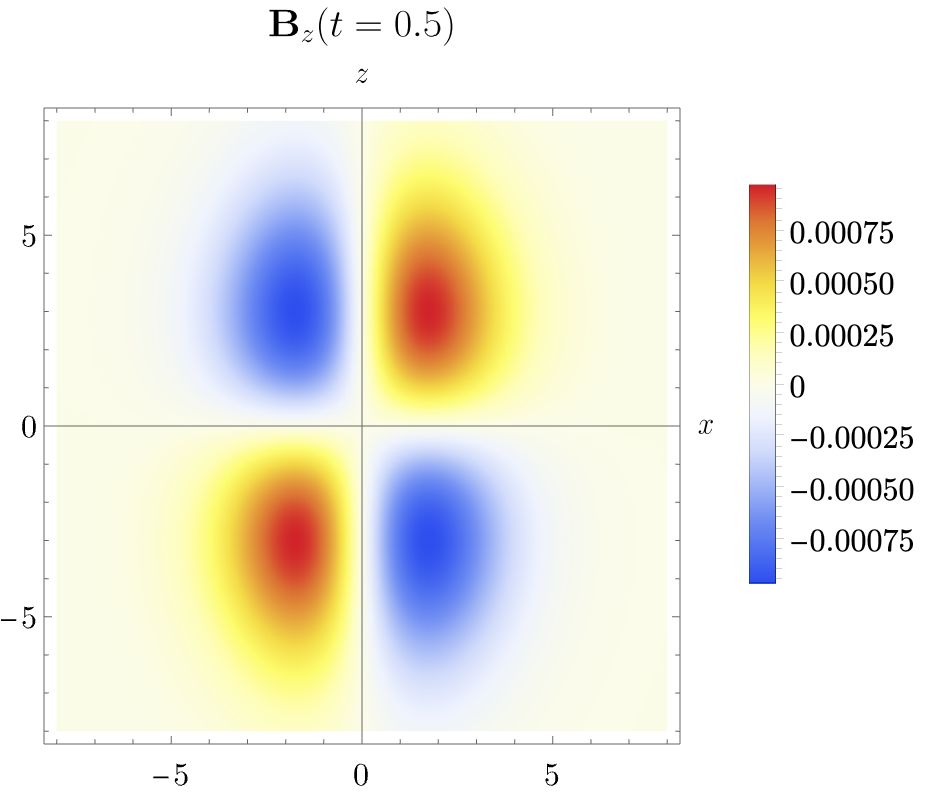}
\includegraphics[width=0.238\textwidth]{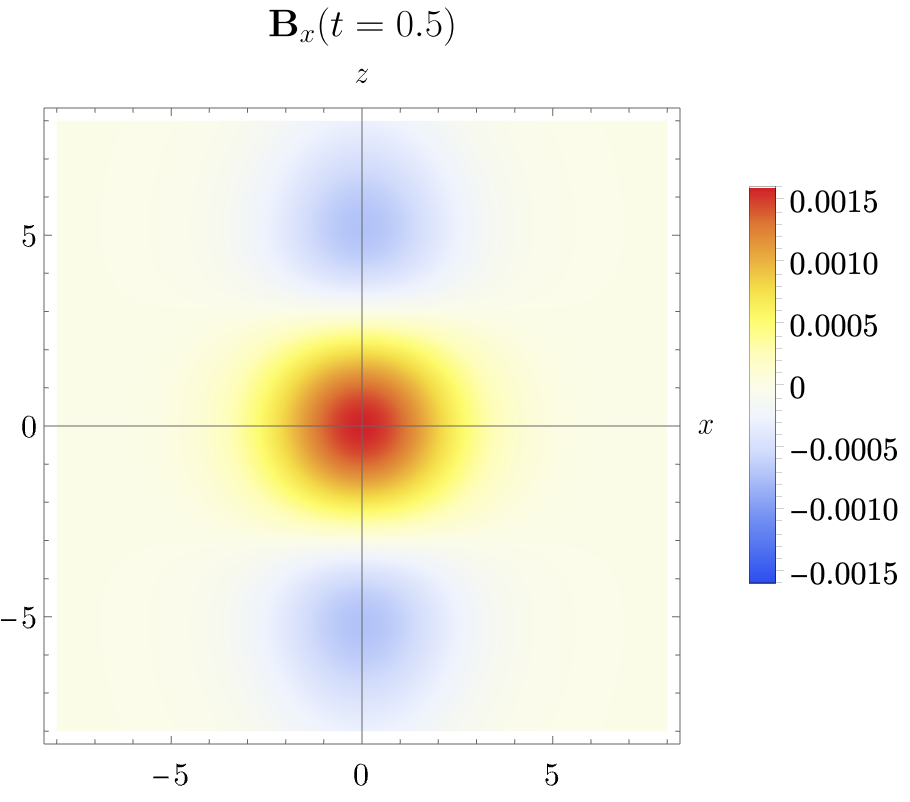}
\caption{Different components of ${\bf B}$ at $t=0.5$, evolved from the initial configuration ${\bf B}(0, \vx) =\frac{1}{\pi \sigma^2}\, e^{-\frac{x^2 + z^2}{\sigma^2}} \hat y$. All planes orthogonal to the $y$ direction are identical. The parameter values are $\sigma_1^\perp=4$, $\sigma^\para = 2$, $\sigma_2^\perp=-1$, $\chi_\perp=1$, and $\chi_\para = 8$. } 
\label{fig:y_with_t5}
\end{figure}

A simple example illustrating the consequences of the full dispersion relation~\eqref{MHD_FullDispRel} is given in 
Fig.~\ref{fig:y_with_t5}, where the initial ${\bf B}$ points in the $y$ direction  and is concentrated around the $y$ axis. The $B_y$ component diffuses anisotropically, and in addition $B_x$ and $B_z$ components are also generated and subsequently diffuse. 
Similar patterns were previously observed in nonlinear simulations of neutron stars in \cite{vigano2} and of magnetized filamentary clouds in \cite{filament}. We see here that these effects may be understood as simple consequences of the dispersion relations~\eqref{MHD_FullDispRel}. Note that if we had set $\sigma^{\perp}_2=0$ for the initial condition in Fig. \ref{fig:y_with_t5}, we would still see anisotropic diffusion of $B_y$, but the $B_x$ and $B_z$ components would remain zero for all times. 

\begin{figure}[ht!] 
\centering
\includegraphics[width=0.3\textwidth]{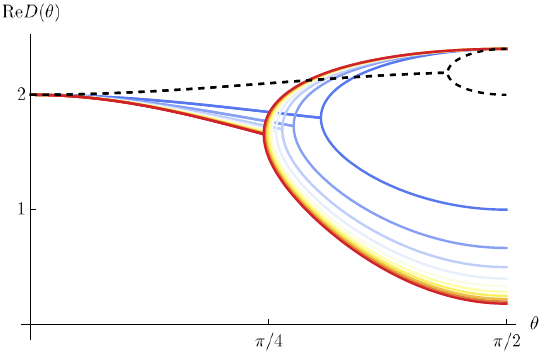}
\includegraphics[width=0.3\textwidth]{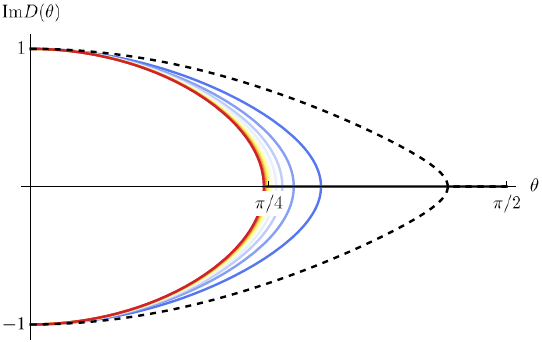}
\caption{Real and imaginary parts of $D(\theta) = i \om/k^2$ for the following choice of parameters: $\beta_0 = \mu = p = a = d = 1$ and $\tilde d = 0.1$, where $k_\perp = k \sin\theta$ and $k_z = k\cos\theta$, for $\theta \in [0,\pi/2]$. The black dashed line represents $\tilde a = 0$ (corresponding to the theory in~\cite{1992ApJ...395..250G}) and the colored lines represent $\tilde a = {1,2,\ldots,10}$, going from blue to red with other parameters fixed.}
\label{fig:DispRel}
\end{figure}

Equation~\eqref{MHD_FullDispRel} has the form $\om = - i D (\theta) k^2$, where $k^2 = k_z^2 + k_\perp^2$ and $\theta$ is the angle between the background $\bf B$ and $\bk$. In general, for $\sig_2^\perp \neq 0$, $D (\theta)$ is complex, with its real part being the ``diffusion constant'' and the imaginary part describing oscillatory propagation. Such propagating behavior was considered in \cite{1992ApJ...395..250G} as a mechanism for the transport of magnetic energy from the inner to the outer crust in neutron stars. $D (\theta)$ at  $\theta =0$ and $\theta = {\pi \ov 2}$ can be read from~\eqref{yeun1}--\eqref{yeun2}, while for general $\theta$, it is plotted in Fig.~\ref{fig:DispRel}. There is a critical value $\theta_c$ beyond which $\text{Im}D(\theta)$ for both modes vanishes, and the real parts become different.  We also notice that as $\chi_\para - \chi_\perp$ is increased (with other parameters fixed), $\theta_c$ decreases.

The effective action formulation allows us to understand the precise way in which the sources $b_{\mu\nu}$ are coupled to 
the magnetohydrodynamic variables, and hence enables us to obtain the retarded two-point functions among components of $\bE$ and $\bB$. The full expressions for generic $\bk$ are complicated, and are presented in SM, Section II. As examples, we show the special cases of all {\em nonzero} correlation functions with $k_z \neq 0$ and $k_x = k_y = 0$ in the Appendix.

{\bf Discussion.}---We have developed a new formulation for MHD applicable to neutron stars, in the presence of strong magnetic fields and in the regime where flow velocities and temperature variations can be neglected. This theory leads to the most general  constitutive relation for the electric field in terms of the magnetic field for this setup, which in particular reproduces the phenomena of Ohmic decay, ampibolar diffusion, and Hall drift derived in previous formulations such as~\cite{1992ApJ...395..250G}. Moreover, the constitutive relation includes new terms which capture implications of a nontrivial magnetic permeability.
As a simple application of our formulation, we examined the behavior of small perturbations on top of a strong constant magnetic field, and uncovered novel magnetic diffusion patterns. We also found the two-point correlation functions among electric and magnetic field components. 

It will be of great interest to apply our formalism to more realistic problems concerning neutron stars and other astrophysical systems with strong magnetic fields, using full nonlinear evolution based on \eqref{ouC}. In particular, it would be interesting to see if the new terms in the second line of \eqref{ouC} lead to qualitatively new phenomena which were not accessible with earlier simulations such as those  in \cite{crust, core, combined}. The MHD formulation here can be generalized to include flow velocities, going beyond the static neutron approximation, as well as temperature variations. It would also be interesting to include the effects of superfluidity or superconductivity, extending the discussion of \cite{10.1111/j.1365-2966.2010.17484.x, dommes}. 

\begin{acknowledgments}
{\bf Acknowledgments.}---We would like to thank Deepto~Chakrabarty, Michael Landry, Nick Poovuttikul and in particular Andreas Reisenegger for very helpful discussions. S.G. was supported by the STFC Ernest Rutherford Fellowship ST/T00388X/1. This work is also supported by the research programme P1-0402 and the project N1-0245 of Slovenian Research Agency (ARIS), and by the Office of High Energy Physics of U.S. Department of Energy under grant Contract Number  DE-SC0012567 and DE-SC0020360 (MIT contract \# 578218). S.L. acknowledges the support of the Natural Sciences and Engineering Research Council of Canada (NSERC). 
\end{acknowledgments}

\begin{appendix}
{\bf Appendix on correlation functions.}---In this appendix, we write the correlation functions involving the ${\bf B}$ and ${\bf E}$ fields for the special case of $\omega \neq 0$, $k_z \neq 0$, and $k_x = k_y = 0$. The complete set of all correlation functions is derived and shown in SM. The {\it nonzero} correlation functions among the ${\bf B}$ components are given by
\begin{align}  
G^R(B_x, B_x) &=\frac{k_z^2 \chi_{\perp}(k_z^2 \xi^2 - i \sigma_1^\perp \chi_{\perp} \omega)}{k_z^4\xi^2 - 2 i k_z^2 \sigma_1^\perp\chi_{\perp} \omega - \chi_{\perp}^2 \omega^2} , \\ \qquad G^R(B_x, B_y) &= \frac{-i k_z^2 \chi_{\perp} \sigma_2^\perp\chi_{\perp}\omega}{k_z^4\xi^2 - 2 i k_z^2 \sigma_1^\perp\chi_{\perp} \omega - \chi_{\perp}^2 \omega^2}   ,
\end{align} 
where $\xi = \sqrt{{\sigma_1^{\perp}}^2 + {\sigma_2^{\perp}}^2}$. The mixed  nonzero correlation functions among ${\bf E}$ and ${\bf B}$ field components are  
\begin{align}  
G^R(B_x, E_x) &= \frac{\omega}{k_z} G^R(B_x, B_y) , \\
G^R(B_x, E_y) &= - \frac{\omega}{k_z} G^R(B_x, B_x)   ,
\end{align}
and among the ${\bf E}$ field components are 
\begin{align} 
G^R(E_z, E_z) &= - \frac{4 k_z^2 m^2 \mu^2}{\chi_{\para}} + i \sigma^\para\omega, \label{e1} \\
 G^R(E_x, E_x) &= \frac{\omega^2}{k_z^2} G^R(B_x, B_x) , \\
 G^R(E_x, E_y) &= \frac{\omega^2}{k_z^2} G^R(B_x, B_y)  .
 \label{y3} 
\end{align} 
\eqref{y1t}--\eqref{y7t} arise by taking $k_z \to 0$ in the above expressions, while the definition of $\chi_{\perp}$ in \eqref{y2t} can be seen by taking $\om \to 0$ with $k_z$ finite. Note that while from \eqref{ouC}, the coefficient $m$ plays no role in the equation of motion of the magnetic field at leading order in derivatives, it does appear at leading order in the correlation functions among electric field components, as in \eqref{e1}.
\end{appendix}

\onecolumngrid

\appendix
\section{SUPPLEMENTARY MATERIAL}

\section{Appendix A: Details of the effective CTP action for MHD}

In this Appendix, we outline the steps leading to Eqs.~(3)--(5) in the Letter. See~\cite{long} for technical details, other generalizations, and a more extensive discussion. 

The hydrodynamical variables for MHD in the static neutron approximation are given by two vector fields $A_{1 \mu}, A_{2 \mu}$, as on a closed time path all quantities come in two copies. The one-form symmetry implies that the action $S_{\rm EFT}$ can only depend on the dynamical variables $A_{s \mu}$ ($s$ labels the copies)
 through the combinations
\begin{equation}\label{BDef}
G_{s\mu\nu} \equiv b_{s\mu\nu} + \partial_\mu A_{s\nu} - \partial_\nu A_{s\mu}  , \quad s =1 ,2 ,
\end{equation}
where  $b_{\mu \nu}$ denotes the external source for the two-form current $J^{\mu \nu}$. 
By definition, $G_{s\mu \nu}$ and thus $S_{\rm EFT}$ are invariant under the transformation $A_{s \mu} \rightarrow A_{s \mu} + \partial_{\mu} \alpha_{s}$, for arbitrary functions $\alpha_s$. We can use the freedom to set $A_{s0} = 0$.  
$S_{\rm EFT} [G_{1\mu \nu}, G_{2\mu \nu}]$ is then the most general action invariant under a number of symmetries. 

To describe the symmetries to be imposed,  it is convenient to introduce the Keldysh basis of symmetric (``$r$'') and antisymmetric (``$a$") combinations of all fields: $\varphi_r \equiv (\varphi_1 + \varphi_2)/2$ and $\varphi_a \equiv \varphi_1 - \varphi_2$. The symmetries that
the MHD action $S_{\rm EFT}$ should satisfy include~\cite{Crossley:2015evo,Glorioso:2017fpd}:
\be\label{dshi}
\text{Diagonal shift}: \quad A_{ri} (t, \vx) \rightarrow A_{ri}(t, \vx) + \lambda_i(\vx) ,
\ee
and the following constraints from unitarity: 
\begin{align}
S_{\rm EFT}[G_{r\mu\nu}, G_{a\mu\nu} = 0] &= 0, \\
S^*_{\rm EFT} [G_{r\mu\nu}, G_{a\mu\nu}] &= - S_{\rm EFT} [G_{r\mu\nu}, - G_{a\mu\nu}], \\ \label{Im}
\text{Im} \, S_{\rm EFT} &\geq 0  .
\end{align}
We assume that the underlying microscopic system has an 
anti-unitary discrete symmetry $\Th$ involving time-reversal $\CT$. The macroscopic implications 
of $\Th$ and the fact that the system is in local equilibrium are realized by 
 a $\mathbb{Z}_2$ dynamical KMS symmetry 
\begin{equation}\label{DynKMS}
\begin{aligned}
 G_{a\mu\nu} (x) & \to  \Theta G_{a\mu\nu}(x) - i \beta_0 \partial_0 \Theta G_{r\mu\nu} (x) , \\  
 G_{r\mu\nu} (x) & \to  \Theta G_{r\mu\nu} (x)   .
\end{aligned}
\end{equation}
Here $\beta_0$ is the inverse temperature. Finally, we need to impose on $S_{\rm EFT}$ any other discrete symmetry respected by 
the underlying system. Depending on what discrete symmetries are present, we obtain different classes of actions, which are fully classified in~\cite{long}.

As discussed in the main text, in neutron stars, the charge conjugation symmetry $\CC$ is badly broken, while parity $\CP$ may be broken due to weak interactions. At the derivative order of the EFT that we work with, parity breaking does not yield any difference in either the dispersion relations or the two-point functions (see~\cite{long} for details). Below, we will first write down the theory with parity conservation and then comment on the parity violating case. The microscopic system respects time reversal $\CT$. With $\CP$ conserved, we can take $\Th$ to be either $\CT$ or $\CP \CT$, which give equivalent actions. The identification in Eq.~(2) in the Letter determines how each discrete symmetry acts on  $J^{\mu \nu}$, and since the source $b_{\mu \nu}$ should transform in the same way as $J^{\mu \nu}$, we conclude that
\begin{align}
\CP :&\quad (G_{0i} , G_{ij} ) \to ( G_{0i} , - G_{ij} ) , \label{P} \\
\CT :&\quad (G_{0i} , G_{ij} ) \to ( - G_{0i} ,  G_{ij} ) , \label{T} \\
\CC:&\quad  (G_{0i} , G_{ij} ) \to ( - G_{0i} , - G_{ij} )  . \label{C} 
\end{align}

Our proposal of MHD for neutron stars is then given by the most general action invariant under~\eqref{dshi}--\eqref{DynKMS} with $\Theta = \CT$ and with invariance under $\CP$.  With $S_{\rm EFT} = \int d^4 x \, \CL_{\rm EFT}$, we write 
\be \label{hen}
\CL_{\rm EFT} = \CL_0  +  \CL_{\CP,  \CT} ,
\ee
where $\CL_0$ contains terms invariant under all the $\CC, \CP, \CT$ symmetries, while $\CL_{\CP , \CT}$ includes 
those violating $\CC$. To first derivative order, $\CL_0$ has the form 
\begin{align}\label{minac}
\CL_0 &= a \,  G_{r0i} G_{a0i} +
 i (c \, \delta_{ij} + \tilde c \, G_{r0i} G_{r0j}  ) \, G_{a0i} \tilde G_{a0j} 
  +i  (d \, \delta_{ik} \delta_{jl}  + \tilde d \, \epsilon_{ijm} {\epsilon}_{kln} G_{r0m} G_{r0n}) \,  G_{aij} 
  \tilde G_{a k l} ,
\end{align}
where $\tilde G_{a \mu \nu} =  G_{a\mu \nu} + i  \beta_0 \pt_0 G_{r\mu \nu}$, and 
$\CL_{\CP, \CT}$ is written as 
\begin{align}
\label{lpt}
 \CL_{\CP , \CT} &= {\epsilon}_{ijk} G_{r0k}  \le(h   \pt_0 G_{r0j} G_{a0i}  +   p  \partial_0 G_{rjl}  G_{ail}\ri)    + \ep_{ijk} \le({f \ov 6} H_{aijk} + f' G_{r0k} H_{rlij} G_{a0l} \ri), 
\end{align}
where $H_{\mu \nu \lambda}  = \partial_{\mu} b_{ \nu \lambda} + \partial_{\lambda} b_{\mu \nu} + \partial_{\nu} b_{\lambda\mu}$. 
Various coefficients $a, c, \tilde c, d, \tilde d, h, p, f$ are functions of $g \equiv G_{r0i}^2$. 
Note that invariance under~\eqref{dshi} implies that $G_{rij}$ must enter $S_{\rm EFT}$ through $\pt_0 G_{rij}$ or $H_{rijk}$, {and note that} 
\begin{align}\label{FieldRel1}
\partial_0 G_{rij} = H_{r0ij} + \partial_i G_{r0j} - \partial_j G_{r0i}  . 
\end{align}

We will see that the system is dominated by diffusion at leading order, which means that when doing derivative counting we should assign weight $2$ to $\p_0$,  weight $1$ to $\p_i$, and weight $0$ to $G_{r0i}$.  Under this assignment, $G_{rij}, G_{a0i}, G_{aij}$ have respective weights $-1, 2, 1$. The terms proportional to $c, \tilde c, h$ then have weights $3$ and are higher order than the other terms which have weights $2$. We will suppress them below.  When parity is broken, additional first derivative terms exist, but they all have higher weights. 

Equations of motion following from $S_{\rm EFT}$ are 
\begin{gather} \label{eom1}
\p_\mu J^{\mu \nu}_r = 0, \quad \p_\mu J^{\mu \nu}_a = 0 , \\
J^{\mu \nu}_r \equiv {\de S_{\rm EFT} \ov \de b_{a \mu \nu }} , \quad J^{\mu \nu}_a \equiv {\de S_{\rm EFT} \ov \de b_{r \mu \nu }}  .
\label{eom2} 
\end{gather} 
With $b_{a \mu \nu} =0$, the second equation  of~\eqref{eom1} can be solved by setting $G_{a\mu \nu} =0$, in which case 
$J^{\mu \nu}_a$ are identically zero. With $G_{\mu \nu} \equiv G_{r \mu \nu} $ and  $J^{\mu\nu} \equiv J_r^{\mu\nu} $, the equations of motion then reduce to Eq.~(2) in the Letter with  $J^{\mu \nu}$ (to weight $1$) given by~(3)--(4). Note that $m$ appearing in~(3)--(4) is equal to $f'$. The first lines of~(3)--(4) come from $\CL_0$ while the second lines come from 
$\CL_{\CP, \CT}$. Equations~(3)--(4) along with~(2) give our new formulation of MHD equations  for neutron stars to first derivative order. 
Lastly, equation~\eqref{Im} requires that
\be 
c \geq0 , \quad 
c + \tilde c \, G_{r0i}^2 \geq 0 , \quad  d \geq 0 , \quad 
d + 2 \tilde d \,  G_{r0i}^2 \geq 0  . 
\ee

\section{Appendix B: Derivation of dispersion relations and linear response theory}

Here we study the linear responses of the system (in the presence of a  constant magnetic field) to an infinitesimal external source. 
From the responses of various components of electric and magnetic fields we can extract physical interpretations for various parameters introduced in Eqs.~(15)--(17) in the Letter. 

Recall that 
\be
J^{\mu\nu} =\tilde F^{\mu\nu} = \frac{1}{2} \epsilon^{\mu\nu\rho\sigma} F_{\rho\sigma}
\ee
and let  $a_\mu$ be the vector potential for $F_{\mu \nu}$. We then have 
\begin{gather}
\ha \int d^4 x \, J^{\mu\nu} b_{\mu\nu} = \int d^4 x \, j^\mu a_\mu, \\
j^\mu = \frac{1}{2} \epsilon^{\mu\nu\rho\sigma} \p_\nu b_{\rho \sigma} 
= {1 \ov 6} \epsilon^{\mu\nu\rho\sigma} H_{\nu\rho\sigma} .
\end{gather}
$j^\mu$ thus corresponds to an external current for $a_\mu$, and
\be 
 j^0 = H_{123}, \qquad j^i = -{1 \over 2}  \ep^{ijk} H_{0jk}  .
\ee
Given the expansions~(12)--(14), the linearized  equations of motion~(2)
can be written as 
can be written as 
\begin{align}
&\tilde K = -{\chi_\para \ov \chi_\perp} K  - \frac{2m\mu}{\chi_\perp} \partial_z H_{xyz} , \\
\label{he1}
&\chi_\para  \p_0 f_{0z} + \sigma_1^{\perp}   ( \p_z \tilde K - \p_a^2 f_{0z})   + \sigma_2^\perp   \p_z K_{xy}   
= -   \sigma_1^{\perp}  \p_a H_{0za} -  \sigma_2^{\perp} \epsilon_{ab} \p_a  H_{0zb}  - 2 m \mu \p_0 H_{xyz}, \\
 \label{he2} 
&\chi_\perp  \p_0 f_{0a} + \sigma^{\para}  \ep_{ab}  \p_b K_{xy}  
 -   \sigma_1^{\perp}  ( \p_z^2 f_{0a} - \p_a K) - \sigma_2^{\perp} \ep_{ab} ( \p_z^2 f_{0b} - \p_b K)  
 = - \sigma^{\para}  \ep_{ab}  \p_b H_{0xy}+    \sigma_1^{\perp} \p_z  H_{0za} + \sigma_2^{\perp} \ep_{ab} \p_z H_{0zb} , 
\end{align}
where we have introduced 
\be 
K \equiv \p_z f_{0z}, \qquad \tilde K \equiv \p_a f_{0a}  , 
\ee
which can be further written as 
\begin{align} \label{DD1}
(\p_0 + \CD ) \bma K \cr K_{xy} \ema = \bma s \cr s_{xy} \ema ,  
\qquad \CD =  \bma D_{11}  & D_{12} \cr D_{21} & D_{22}  \ema ,
\end{align}
where 
\begin{align}
D_{11} &=  - \sig^{\perp}_1 \le[ {1 \ov \chi_\perp} \p_z^2 + {1 \ov \chi_\para} \p_a^2 \ri]
=   \sig^{\perp}_1 \le[ {1 \ov \chi_\perp} k_z^2 + {1 \ov \chi_\para} k_a^2 \ri] , \\
D_{12} &=  {\sig_2^\perp \ov \chi_\para}
 \p_z^2  = - {\sig_2^\perp \ov\chi_\para} k_z^2 , \\
D_{21} &=  -{ \sig_2^{\perp}  \chi_\para  \ov  \chi_\perp} \le({\p_z^2 \ov \chi_\perp} + {\p_a^2 \ov \chi_\para} \ri)   ={ \sig_2^{\perp}  \chi_\para  \ov  \chi_\perp} \le({k_z^2 \ov \chi_\perp} + {k_a^2 \ov \chi_\para} \ri) , \\
D_{22} &= - {1 \ov \chi_\perp} \le[ \sig^\perp_1  \p_z^2  + \sigma^{\para} \p_b^2 \ri] = 
{1 \ov \chi_\perp} \le[ \sig^\perp_1  k_z^2  + \sigma^{\para} k_b^2 \ri] , 
\end{align}
and
\begin{align}
s &= -   {\sigma_1^{\perp} \ov \chi_\para } \p_z  \p_a H_{0za} - {\sigma_2^{\perp}  \ov \chi_\para } \epsilon_{ab} \p_a \p_z H_{0zb} -\frac{2m \mu}{\chi_{\para}}(\partial_0 \partial_z H_{xyz} - \frac{\sigma_1^{\perp}}{\chi_{\perp}}\partial_z^3 H_{xyz}), \\
s_{xy} &=  {\sigma^{\para} \ov  \chi_\perp }  \p_a^2 H_{0xy}+   { \sigma_1^{\perp} \ov  \chi_\perp } \epsilon_{ab} \p_a \p_z H_{0zb}-  {\sigma_2^{\perp}  \ov  \chi_\perp } \p_z \p_a H_{0za} + \frac{2 m \mu \sigma_2^{\perp}}{\chi_{\perp}^2} \partial_z^3 H_{xyz} .
\end{align}

In momentum space, we then have 
\bea 
 \bma K \cr K_{xy} \ema &=&  {1 \ov - i \om + \CD } \bma s \cr s_{xy} \ema =  {1 \ov \Lambda} \bma - i \om + D_{22} & - D_{12} \cr - D_{21} & - i \om + D_{11} \ema   \bma s \cr s_{xy} \ema ,
\eea
where $\Lambda = \det (- i \om +\CD)$. It will also be useful to introduce in some of the expressions below the quantity 
\be 
\Gamma = \chi_{\para} \chi_{\perp}^2 \Lambda  . 
\ee

Let us first set the external sources to zero, so that all $H_{\mu \nu \sigma}$ and hence $s$, $s_{xy}$ are zero. We must then have $\text{det} \Lambda = 0$, which leads to the dispersion relations~(23) in the Letter. 

In general, the $s$, $s_{xy}$ are non-zero, and we can solve for $f_{0i}$ in terms of these sources by noting that 
\begin{align}  
f_{0z} &= - {i \ov k_z} K, \\ 
f_{0a} &= {i \ov k_\perp^2} (k_a {\chi_\para \ov \chi_\perp} K + \ep_{ab} k_b K_{xy}) - \frac{2m \mu}{\chi_{\perp} k_{\perp}^2} k_x k_z H_{xyz} ,
\end{align} 
which gives 
\begin{align} 
f_{0z} &=  - {i \ov k_z \Lam} \le((- i \om + D_{22}) s - D_{12} s_{xy} \ri) , \\
f_{0a} &= {i \ov k_\perp^2 \Lam} \le[k_a {\chi_\para \ov \chi_\perp}  ((- i \om + D_{22}) s - D_{12} s_{xy} )  + \ep_{ab} k_b ( - D_{21}s + (- i \om + D_{11}) s_{xy} ) \ri]   - \frac{2m \mu}{\chi_{\perp} k_{\perp}^2} k_x k_z H_{xyz}  .
\end{align}
These expressions can be substituted back into Eqs.~(12)--(14) in the Letter to find expressions for $J^{\mu \nu}$ in terms of the external sources $b_{\mu \nu}$, which in turn can be used to find the retarded correlation functions 
\begin{gather} 
G^R(J^{\mu \nu} , J^{\lambda \rho};  t_2 - t_1, {\bf x}_2 - {\bf x}_1)  \equiv i \theta(t_2 - t_1) \langle [J^{\mu \nu}(t_2, {\bf x_2}),  J^{\lambda \rho }(t_1, {\bf x_1})] \rangle = \frac{\delta J^{\mu \nu}(t_2)}{\delta b_{\lambda \rho}(t_1)} \, . 
\end{gather}
(See for instance \cite{Crossley:2015evo} for details on why the last two expressions above are equal.)
From the invariance of the above expression under $\CP \CT$ symmetry,  the retarded functions should satisfy 
\be \label{sym1}
G^R(J^{\mu \nu} , J^{\lambda \rho}) = G^R(J^{\lambda \rho}, J^{\mu \nu})_{\mu \rightarrow - \mu}  . 
\ee
From $J^{\mu \nu}$ being Hermitian, we have 
\be 
{G^R(J^{\mu \nu} , J^{\lambda \rho}; k^{\sigma})}^{\ast} = G^R(J^{\mu \nu} , J^{\lambda \rho}; - k^{\sigma})  
=  \pm G^R(J^{\mu \nu} , J^{\lambda \rho} ; \, - \om, \vk) ,
\ee
where in the last expression we have also used parity symmetry. 
Further constraints on the correlation functions are given by Ward identities 
\be 
k_\mu G(J^{\mu \nu} , J^{\lambda \rho}; k^{\sigma}) = 0, \label{sym2}
\ee
and, moreover, there is a rotational symmetry in the $x-y$ plane. 

The expressions for the retarded correlation functions with general $k_i$, $\omega$, and $\mu$ are complicated, so let us first consider some special limits, which in particular will justify the interpretations of various coefficients given in Eqs.~(18) and (19)--(21) in the Letter. 

\subsection{Case 1: $\mu =0$} 

In this case, the system becomes rotationally symmetric and we have $\sigma_1^{\perp}=\sigma^{\para}=\sigma$, $\chi_{\perp}=\chi_{\para}=\chi$, and 
\be 
\Ga = \chi (\sig k^2 - i \chi \om)^2 .
\ee
The correlation functions are given by 
\begin{gather} 
G(B_i,  B_j) = \frac{ (k^2 \delta_{ij} - k_i k_j) \sigma \chi}{k^2 \sigma -i \chi \omega } ,
\qquad G(B_i, E_j) = \frac{- \epsilon_{ijm} k_m \sigma \chi \omega}{k^2 \sigma -i \chi \omega}, 
 \\ 
 G(E_i, E_j) = \frac{\sigma \omega(i k_i k_j \sigma + \delta_{ij} \chi \omega)}{k^2 \sigma -i \chi \omega}
 =i  \sig \om \de_{ij} +i \sigma^2 \omega  \frac{k_i k_j -  k^2 \de_{ij} }{k^2 \sigma -i \chi \omega}  .
\end{gather} 
Any correlation functions that are not written explicitly can be derived using the symmetries listed around \eqref{sym1}--\eqref{sym2}
Note further that the expression for $G (B_i, B_j)$ can be determined from rotational symmetry and $\nab \cdot {\bf B} =0$. Similarly with $G(B_i, E_j)$ and $G(E_i, E_j)$.

By rotational symmetry we can align momentum to be along the $z$ direction and the above expressions can be written more explicitly as 
\bega \label{henl}
G^R( B_z, B_i) =0, \qquad G^R(B_a, B_b) = {\sigma \chi k_z^2 \de_{ab}  \ov k^2_z \sigma -i \chi \omega }, 
\qquad G^R(E_z, E_z) = i \sig \om , \qquad G^R(E_z, E_{a}) =0 ,  \\
G^R(E_a, E_b) = i  \sig \om \de_{ab} - i \sigma^2 \omega  \frac{ k^2_z \de_{ab} }{k_z^2 \sigma -i \chi \omega}
=  i  \sig \om \de_{ab} - i {\sig \om \ov \chi} G^R(B_a, B_b)  . 
\end{gather} 
We can then interpret $\sig$ as a magnetic ``viscosity'' as defined by the Kubo formula 
\be \label{y1}
\sig = \lim_{\om \to 0} \lim_{k_z \to 0} {1 \ov i \om} G^R(E_z, E_z) 
\ee
and $\chi$ as the one-form susceptibility defined by 
\be \label{y2}
\chi = \lim_{k_z \to 0} \lim_{\om \to 0}  G^R(B_z, B_z)  .
\ee
We note that the order of limits are important in~\eqref{y1}--\eqref{y2}.

\subsection{Case 2: $\mu \neq 0$ and momentum along the $z$ direction}

In this case we have 
\be 
\Ga = k_z^4\xi^2 - 2 i k_z^2 \sigma_1^\perp\chi_{\perp} \omega - \chi_{\perp}^2 \omega^2
=  (k_z^2 \sigma_2^\perp)^2 + ( k_z^2 \sigma_1^\perp - i \chi_{\perp} \omega)^2 ,
\ee
where 
\be 
\xi = \sqrt{{\sigma_1^{\perp}}^2 + {\sigma_2^{\perp}}^2} . 
\ee
The correlation functions of the magnetic fields are 
\begin{gather}  
G^R(B_z, B_a) = 0, \qquad G^R(B_x, B_x) =\frac{k_z^2 \chi_{\perp}(k_z^2 \xi^2 - i \sigma_1^\perp \chi_{\perp} \omega)}{k_z^4\xi^2 - 2 i k_z^2 \sigma_1^\perp\chi_{\perp} \omega - \chi_{\perp}^2 \omega^2} , \\ \qquad G^R(B_x, B_y)= \frac{-i k_z^2 \chi_{\perp} \sigma_2^\perp\chi_{\perp}\omega}{k_z^4\xi^2 - 2 i k_z^2 \sigma_1^\perp\chi_{\perp} \omega - \chi_{\perp}^2 \omega^2}   .
\end{gather} 
The mixed correlation functions among electric and magnetic fields are 
\begin{gather}  
G^R(B_z, E_a) = G^R(B_x, E_z) = 0 , \qquad    
G^R(B_x, E_x) = \frac{\omega}{k_z} G^R(B_x, B_y) , \qquad
G^R(B_x, E_y) = - \frac{\omega}{k_z} G^R(B_x, B_x)   .
\end{gather}
The correlation functions among electric field components are 
\begin{gather} 
G^R(E_z, E_z) = - \frac{4 k_z^2 m^2 \mu^2}{\chi_{\para}} + i \sigma^\para\omega, \qquad G^R(E_z, E_a) = 0 , \\
 G^R(E_x, E_x) = \frac{\omega^2}{k_z^2} G^R(B_x, B_x) , \qquad 
 G^R(E_x, E_y) = \frac{\omega^2}{k_z^2} G^R(B_x, B_y)  .
 \label{y3} 
\end{gather}

Now taking $k_z \to 0$ the only non-vanishing pieces are 
\be \label{vas}
G^R(E_z, E_z) = i \sigma^\para\omega,  \qquad  G^R(E_x, E_x) = i \sigma_1^\perp \omega, \qquad G^R(E_x, E_y) = i \sigma_2^\perp \omega  . 
\ee
This leads to Eqs.~(19)--(21) in the Letter, and we now interpret $\sig^{\para}, \sigma_1^\perp$ and $\sigma_2^\perp$ respectively as the longitudinal, transverse and Hall magnetic viscosities.  

With $\om \to 0$ (and $k_z$ finite) we have 
\be 
G^R(B_x, B_x) = \chi_\perp,  \qquad G^R(E_z, E_z) = - \frac{4 k_z^2 m^2 \mu^2}{\chi_{\para}}  .
\ee
This gives the second equation in (18), and we thus should interpret $\chi_\perp$ as transverse one-form susceptibility.

\subsection{Case 3: $\mu \neq 0$ and momentum along transverse direction} 

For the spatial momentum along $k_\perp$ we have 
\be 
\Ga =   \chi_{\perp}(\sigma_1^\perp k_{\perp}^2 - i \chi_{\para} \omega)(\sigma^\para k_{\perp}^2 - i \chi_{\perp} \omega)  . 
\ee
The correlation functions among magnetic field components are 
\begin{gather}  
G^R(B_z, B_z) = \frac{\chi_{\para} \sigma_1^\perp k_{\perp}^2}{k_{\perp}^2 \sigma_1^\perp - i \chi_{\para}\omega}, \qquad G^R(B_x, B_x) =  \sigma^\para \chi_{\perp} \frac{k_y^2}{k_{\perp}^2\sigma^\para - i \chi_{\perp}\omega}, \\ 
G^R(B_z, B_x) = 0, \qquad G^R(B_x, B_y)= - \chi_{\perp} \sigma^\para \frac{k_x k_y}{k_{\perp}^2 \sigma^\para - i \chi_{\perp}\omega} .
\end{gather} 
The mixed correlation functions among electric and magnetic fields are 
\begin{gather}  
G^R(B_z, E_x) = \frac{-2 i m \mu k_x \sigma_1^\perp k_{\perp}^2 + \chi_{\para}\omega(- k_y \sigma_1^\perp - k_x \sigma_2^\perp) }{\sigma_1^\perp k_{\perp}^2 - i \chi_{\para}\omega} , \\  
G^R(B_z, E_z) = G^R(B_x, E_x) =G^R(B_x, E_y)=0 ,
\qquad G^R(B_x, E_z) = {\om \ov k_y} G^R(B_x, B_x).
\end{gather}
The correlation functions among electric field components are 
\begin{gather} 
G^R(E_z, E_z) = \frac{\sigma^\para\chi_{\perp} \omega^2}{\sigma^\para k_{\perp}^2 -i \chi_{\perp}\omega} = {\om^2 \ov k_y^2} G^R(B_x, B_x) 
, \qquad G^R(E_z, E_x) = 0 , \\
 \qquad G^R(E_x, E_x) = i\omega \frac{ k_x^2({\sigma_1^\perp}^2 +( 2 m \mu- \sigma_2^\perp)^2) - i \omega \sigma_1^\perp \chi_{\para}}{\sigma_1^\perp k_{\perp}^2 - i \chi_{\para}\omega} , \\
 G^R(E_x, E_y) = i \omega \frac{2 m \mu \sigma_1^\perp k_{\perp}^2 + k_x k_y({\sigma_1^\perp}^2+ ( 2 m \mu- \sigma_2^\perp)^2) - i \omega \sigma_2^\perp\chi_{\para}}{\sigma_1^\perp k_{\perp}^2 - i\omega \chi_{\para}} .
\end{gather} 
We can use rotational symmetry to take $k_\perp = k_y$.  We then find that 
\begin{gather}  
G^R(B_z, E_x) = - \frac{ \chi_{\para}\sigma_1^\perp \omega k_y }{\sigma_1^\perp k_{y}^2 - i \chi_{\para}\omega} , 
\qquad G^R(B_z, E_y) = k_y \frac{-2 i m \mu \sigma_1^\perp  k_y^2 - \chi_{\para} \sigma_2^\perp\omega   }{\sigma_1^\perp k_{y}^2 - i \chi_{\para}\omega}, \\
G^R(E_y, E_y) = i\sigma_1^\perp  \omega +  \frac{ k_y^2 (2 m \mu- \sigma_2^\perp)^2 }{\sigma_1^\perp k_{y}^2 - i \chi_{\para}\omega} , \qquad G^R(E_x, E_x) =  \frac{ \omega^2 \sigma_1^\perp \chi_{\para}}{\sigma_1^\perp k_{y}^2 - i \chi_{\para}\omega} , 
\\
 G^R(E_x, E_y) =  i\sigma_2^\perp  \omega + i \sigma_1^\perp k_{y}^2 \omega \frac{2 m \mu   +  \sigma_2^{\perp}   }{\sigma_1^\perp k_{y}^2 - i\omega \chi_{\para}}  .
\end{gather}
A curious thing is that even in the direction parallel to $k_\perp = k_y$, $G^R(E_y, E_y)$ still has a 
 diffusion pole.

With $k_\perp \to 0$, we again find~\eqref{vas}.  For $\om \to 0$, we have 
\be 
G^R(B_z, E_x) = -2 i m \mu k_x , \qquad G^R(B_z,B_z) = \chi_\para ,
\ee
which gives the first equation in (18), and implies that we  should interpret $\chi_\para$ as the longitudinal one-form susceptibility.

\subsection{Case 4: $\mu \neq 0$ and general momentum} 

For the correlation functions among the magnetic field components in the case of general $k$ and $\omega$, it is useful to define the quantity
\be
A=  k_z^2 {\xi}^2 + \sigma_1^{\perp} \sigma^{\para} k_{\perp}^2, 
\ee
Note that we can write
\begin{align}
\Gamma 
 &= \chi_{\para}  k_z^2 (A- i \sigma_1^\perp \chi_{\perp} \omega)  + \chi_{\perp} k_{\perp}^2 (A - i \sigma^\para \chi_{\para} 
\omega) - i  \chi_{\perp}  \sigma_1^\perp  \omega (\chi_{\para} k_z^2 + \chi_{\perp} k_{\perp}^2)
 -   \chi_{\perp}^2 \chi_{\para} \omega^2 \nn 
 &= (\chi_{\para}  k_z^2 + \chi_{\perp} k_{\perp}^2 ) (A- i \sigma_1^\perp \chi_{\perp} \omega)  
 - i \chi_{\para}  \chi_{\perp}  \om (\sigma_1^\perp  k_z^2 + \sigma^\para k_\perp^2)
 -   \chi_{\perp}^2 \chi_{\para} \omega^2 .
\end{align}
We introduce the following combinations of the  $B_a$ and $E_a$ 
\be 
B_1 = {B_a k_a \ov k_\perp} , \qquad B_2 = {\ep_{ab} B_a k_b \ov  k_\perp} , \qquad E_1 = {E_a k_a \ov k_\perp} , \qquad E_2 = {\ep_{ab} E_a k_b \ov k_\perp}  . 
\ee
The retarded correlation functions among magnetic field components are then given by 
\begin{align}
G^R(B^z,  B^z ) &= \frac{\chi_{\para}\chi_{\perp}}{\Gamma} k_{\perp}^2 (A- i \sigma_1^\perp \chi_{\perp} \omega) , \\
G^R(B^z,B_1) &= - \frac{\chi_{\para} \chi_{\perp}}{\Gamma} k_z  k_\perp (A- i \sigma_1^\perp \chi_{\perp} \omega) , \\
G^R(B^z,B_2) &= -i \frac{\chi_{\para} \chi_{\perp}^2 \sigma_2^\perp }{\Gamma} k_z  k_\perp   \omega ,\\ 
G^R(B_1, B_1) &= \frac{ \chi_{\perp} \chi_\para}{\Gamma}  k_z^2 \le( A - i  \sigma_1^\perp  \chi_{\perp}  \om
 \ri),\\
G^R(B_1, B_2) &= i \frac{\chi_{\perp}^2 \chi_{\para} \sigma_2^\perp}{\Gamma}\omega k_z^2 , \\
G^R(B_2 , B_2 ) &= \frac{ \chi_{\perp}}{\Gamma} \left[ \chi_\perp k_\perp^2 (A- i \chi_{\para} \sigma^{\para}  \omega)
+  \chi_\para k_z^2 (A   - i   \chi_{\perp}  \sigma_1^\perp  \omega  )  \right].
\end{align}

Among the electric field components we have: 
\begin{align}
G^R(E^z, E^z) = & \, i \omega \sigma^\para - \frac{4 m^2 \mu^2}{\chi_{\para}}k_z^2 + \frac{4 m^2 \mu^2 \chi_{\perp}}{\chi_{\para}}\frac{k_z^2 k_{\perp}^2}{\Gamma}(A - i\sigma_1^\perp\chi_{\perp}\omega) \cr & - i \frac{\omega k_{\perp}^2 \sigma^\para}{\Gamma} \left[  k_z^2(\sigma_1^\perp\sigma^\para\chi_{\para} + 4 m \mu \sigma_2^{\perp} \chi_{\perp})+\sigma^\para\chi_{\perp}(\sigma_1^\perp k_{\perp}^2-i \chi_{\para}\omega) \right]  ,
\end{align}
\begin{align}
G^R(E^z, E_1) =& \, \frac{k_\perp k_z}{\Gamma} \biggr[ k_z^2 \left(i  \omega (\xi^2 \sigma^\para \chi_{\para} + 8 m^2 \mu^2 \sigma_1^{\perp} \chi_{\perp}) - 4 m^2 \mu^2 A\right) \nn 
&+ \omega \chi_\perp  \left( i k_\perp^2 ( {\sigma_1^{\perp}}^2 + (2 m \mu - \sigma_2^{\perp})^2) \sigma^\para + \omega  \sigma^\perp_1 \sigma^\para \chi_\para + 2 m \mu \omega \chi_\perp ( 2 m \mu - \sigma_2^\perp) \right)  \biggr],
\end{align}
\begin{align}
G^R(E^z, E_2) =& \, \frac{ k_\perp k_z \omega \chi_\perp}{\Gamma} \left[ -(\sigma_2^{\perp}\sigma^\para\chi_{\para}+2m \mu \sigma_1^{\perp}\chi_{\perp})\,  \omega - 2 i m \mu A \right], 
\end{align}
\begin{align}
G^R(E_1,E_1) 
=& \,\, i \omega \sigma_1^\perp - i \frac{k_\perp^2 k_z^2 \omega}{\Gamma}\left[-(\sigma_2^{\perp})^2 \sigma^{\para} \chi_{\para} +\sigma_1^{\perp}\chi_{\perp}({\xi}^2 - 8 m^2 \mu^2)\right] - \frac{4 m^2 \mu^2 k_\perp^2 k_z^2 A}{\Gamma}  \nn
& - \frac{\omega}{\Gamma} \left[ i k_z^4 \sigma_1^\perp {\xi}^2 \chi_\para + k_z^2 \omega \left({\sigma_1^\perp}^2 - {\sigma_2^\perp}^2\right) \chi_\para \chi_\perp - i k_\perp^2 \chi_\perp (2m\mu- \sigma_2^\perp)^2 \left( k_\perp^2 \sigma^\para - i \omega \chi_\perp   \right)  \right] ,
\end{align} 
\begin{align}
G^R(E_1,E_2) 
=&  -i \omega \sigma^\perp_2 - \frac{i\omega}{\Gamma} \Big[ -k_z^4 \sigma_2^\perp {\xi}^2 \chi_\para + k_\perp^2 \sigma_1^\perp \chi_\perp (2m\mu - \sigma_2^\perp) \left(k_\perp^2 \sigma^\para - i \omega \chi_\perp \right)   \nn
& + k_z^2 \left( 2 i \omega \sigma_1^\perp \sigma_2^\perp \chi_\para \chi_\perp + k_\perp^2 \left( -\sigma_1^\perp \sigma_2^\perp \sigma^{\para} \chi_\para + (2m\mu-\sigma_2^\perp ) {\xi}^2 \chi_\perp \right)   \right) \Big] ,
\end{align}
\begin{align}
G^R(E_2,E_2) 
=& \,\, i \omega \sigma_1^\perp - \frac{i\omega}{\Gamma} \Big[    k_z^4 \sigma_1^\perp {\xi}^2 \chi_\para + k_\perp^2 {\sigma_1^\perp}^2 \chi_\perp \left(k_\perp^2 \sigma^\para - i \omega \chi_\perp\right)  \nn
&+ k_z^2 \left( - i \omega \left( {\sigma_1^\perp}^2 - {\sigma_2^\perp}^2   \right)   \chi_\para \chi_\perp + k_\perp^2 \sigma_1^\perp \left( \sigma_1^\perp \sigma^\para \chi_\para + {\xi}^2 \chi_\perp \right) \right) \Big] .
\end{align}

The mixed retarded correlation functions between the electric and magnetic field components are 
\begin{align}
G^R(B^z, E^z) = \frac{k_\perp^2 k_z \chi_\perp}{\Gamma} \left[ -\omega \sigma_2^\perp \sigma^\para \chi_\para - 2 i m \mu  \left( A - i   \sigma_1^\perp\chi_\perp \omega \right)  \right],
\end{align}
\begin{align}
G^R(B^z,E_1) =- \frac{ i k_\perp \chi_\perp}{\Gamma} \left[ i k_\perp^2 \left( -(\sigma_2^{\perp}\sigma^\para\chi_{\para}+2m \mu \sigma_1^{\perp}\chi_{\perp})\, \omega - 2 m \mu i k_z^2 \xi^2 \right) + 2 m \mu k_\perp^4 \sigma_1^\perp\sigma^\para - \omega^2 \sigma_2^\perp \chi_\para \chi_\perp \right] ,
\end{align}
\begin{align}
G^R(B^z, E_2 ) = \frac{k_\perp \omega \chi_\para \chi_{\perp}}{\Gamma} \left( i \omega \sigma_1^\perp\chi_\perp - k_\perp^2 \sigma_1^\perp\sigma^\para - k_z^2 \xi^2 \right) ,
\end{align}
\begin{align}
G^R(B_1, E^z) = \frac{k_\perp k_z^2 \chi_\perp}{\Gamma} \left[ 2 i m \mu  \left( k_\perp^2 \sigma_1^\perp\sigma^\para + k_z^2 \xi^2 - i \sigma_1^\perp\chi_\perp \omega \right) + \sigma_2^\perp \sigma^\para \chi_\para \omega \right],
\end{align}
\begin{align}
G^R(B_1,E_1) = \frac{i k_z \chi_\perp}{\Gamma} \left[i k_\perp^2 \left( -(\sigma_2^{\perp}\sigma^\para\chi_{\para}+ 2m \mu \sigma_1^{\perp}\chi_{\perp}) \omega - 2 m \mu i  k_z^2 \xi^2 \right) + 2 m \mu k_\perp^4 \sigma_1^\perp\sigma^\para - \omega^2 \sigma_2^\perp \chi_\para \chi_\perp  \right] ,  
\end{align}
\begin{align}
G^R(B_1,E_2 ) = \frac{k_z \omega \chi_\para \chi_\perp}{\Gamma} \left(k_\perp^2 \sigma_1^\perp\sigma^\para + k_z^2 \xi^2 - i \omega \sigma_1^\perp\chi_\perp  \right),
\end{align}
\begin{align}
G^R(B_2,E^z) = \frac{k_\perp \omega \chi_\perp}{\Gamma} \left[ \sigma^\para \chi_\perp \left( k_\perp^2 \sigma_1^\perp- i \omega \chi_\para \right) + k_z^2 \left( \sigma^\perp_1 \sigma^\para \chi_\para + 2 m \mu \sigma_2^\perp \chi_\perp\right)  \right] ,
\end{align}
\begin{align}
G^R(B_2,E_1) =  - \frac{k_z \omega \chi_\perp}{\Gamma} \left[ k_\perp^2 \chi_\perp \left( \xi^2 - 2 m \mu \sigma_2^{\perp} \right) + k_z^2 \xi^2 \chi_\para - i \omega \sigma_1^\perp\chi_\para \chi_\perp \right] ,
\end{align}
\begin{align}
G^R(B_2,E_2) = -\frac{ i k_z \omega^2 \sigma_2^\perp \chi_\para \chi_\perp^2}{\Gamma} .
\end{align}

\bibliography{biblio}{}

\begin{thebibliography}{50}%
\makeatletter
\providecommand \@ifxundefined [1]{%
 \@ifx{#1\undefined}
}%
\providecommand \@ifnum [1]{%
 \ifnum #1\expandafter \@firstoftwo
 \else \expandafter \@secondoftwo
 \fi
}%
\providecommand \@ifx [1]{%
 \ifx #1\expandafter \@firstoftwo
 \else \expandafter \@secondoftwo
 \fi
}%
\providecommand \natexlab [1]{#1}%
\providecommand \enquote  [1]{``#1''}%
\providecommand \bibnamefont  [1]{#1}%
\providecommand \bibfnamefont [1]{#1}%
\providecommand \citenamefont [1]{#1}%
\providecommand \href@noop [0]{\@secondoftwo}%
\providecommand \href [0]{\begingroup \@sanitize@url \@href}%
\providecommand \@href[1]{\@@startlink{#1}\@@href}%
\providecommand \@@href[1]{\endgroup#1\@@endlink}%
\providecommand \@sanitize@url [0]{\catcode `\\12\catcode `\$12\catcode
  `\&12\catcode `\#12\catcode `\^12\catcode `\_12\catcode `\%12\relax}%
\providecommand \@@startlink[1]{}%
\providecommand \@@endlink[0]{}%
\providecommand \url  [0]{\begingroup\@sanitize@url \@url }%
\providecommand \@url [1]{\endgroup\@href {#1}{\urlprefix }}%
\providecommand \urlprefix  [0]{URL }%
\providecommand \Eprint [0]{\href }%
\providecommand \doibase [0]{http://dx.doi.org/}%
\providecommand \selectlanguage [0]{\@gobble}%
\providecommand \bibinfo  [0]{\@secondoftwo}%
\providecommand \bibfield  [0]{\@secondoftwo}%
\providecommand \translation [1]{[#1]}%
\providecommand \BibitemOpen [0]{}%
\providecommand \bibitemStop [0]{}%
\providecommand \bibitemNoStop [0]{.\EOS\space}%
\providecommand \EOS [0]{\spacefactor3000\relax}%
\providecommand \BibitemShut  [1]{\csname bibitem#1\endcsname}%
\let\auto@bib@innerbib\@empty
\bibitem [{\citenamefont {Lattimer}\ and\ \citenamefont
  {Prakash}(2004)}]{Lattimer:2004pg}%
  \BibitemOpen
  \bibfield  {author} {\bibinfo {author} {\bibfnamefont {J.~M.}\ \bibnamefont
  {Lattimer}}\ and\ \bibinfo {author} {\bibfnamefont {M.}~\bibnamefont
  {Prakash}},\ }\bibfield  {title} {\enquote {\bibinfo {title} {{The physics of
  neutron stars}},}\ }\href {\doibase 10.1126/science.1090720} {\bibfield
  {journal} {\bibinfo  {journal} {Science}\ }\textbf {\bibinfo {volume}
  {304}},\ \bibinfo {pages} {536--542} (\bibinfo {year} {2004})},\ \Eprint
  {http://arxiv.org/abs/astro-ph/0405262} {arXiv:astro-ph/0405262} \BibitemShut
  {NoStop}%
\bibitem [{\citenamefont {Kaspi}\ and\ \citenamefont
  {Beloborodov}(2017)}]{Kaspi:2017fwg}%
  \BibitemOpen
  \bibfield  {author} {\bibinfo {author} {\bibfnamefont {Victoria~M.}\
  \bibnamefont {Kaspi}}\ and\ \bibinfo {author} {\bibfnamefont {Andrei}\
  \bibnamefont {Beloborodov}},\ }\bibfield  {title} {\enquote {\bibinfo {title}
  {{Magnetars}},}\ }\href {\doibase 10.1146/annurev-astro-081915-023329}
  {\bibfield  {journal} {\bibinfo  {journal} {Ann. Rev. Astron. Astrophys.}\
  }\textbf {\bibinfo {volume} {55}},\ \bibinfo {pages} {261--301} (\bibinfo
  {year} {2017})},\ \Eprint {http://arxiv.org/abs/1703.00068} {arXiv:1703.00068
  [astro-ph.HE]} \BibitemShut {NoStop}%
\bibitem [{\citenamefont {Harding}\ and\ \citenamefont
  {Lai}(2006)}]{harding2006physics}%
  \BibitemOpen
  \bibfield  {author} {\bibinfo {author} {\bibfnamefont {Alice~K}\ \bibnamefont
  {Harding}}\ and\ \bibinfo {author} {\bibfnamefont {Dong}\ \bibnamefont
  {Lai}},\ }\bibfield  {title} {\enquote {\bibinfo {title} {Physics of strongly
  magnetized neutron stars},}\ }\href@noop {} {\bibfield  {journal} {\bibinfo
  {journal} {Reports on Progress in Physics}\ }\textbf {\bibinfo {volume}
  {69}},\ \bibinfo {pages} {2631} (\bibinfo {year} {2006})}\BibitemShut
  {NoStop}%
\bibitem [{Note1()}]{Note1}%
  \BibitemOpen
  \bibinfo {note} {We use $\mu , \nu , \protect \cdots $ to denote spacetime
  indices and $i,j, \protect \cdots $ to denote spatial indices.}\BibitemShut
  {Stop}%
\bibitem [{Note2()}]{Note2}%
  \BibitemOpen
  \bibinfo {note} {This amounts to assuming that the neutron background is
  static. This assumption has been relaxed in recent works (see e.g.~\cite
  {ofengeim, castillovelocity}), leading to faster evolution of the magnetic
  field, but it can still be used as a first approximation to illustrate the
  main physical effects.}\BibitemShut {Stop}%
\bibitem [{\citenamefont {{Goldreich}}\ and\ \citenamefont
  {{Reisenegger}}(1992)}]{1992ApJ...395..250G}%
  \BibitemOpen
  \bibfield  {author} {\bibinfo {author} {\bibfnamefont {Peter}\ \bibnamefont
  {{Goldreich}}}\ and\ \bibinfo {author} {\bibfnamefont {Andreas}\ \bibnamefont
  {{Reisenegger}}},\ }\bibfield  {title} {\enquote {\bibinfo {title} {{Magnetic
  Field Decay in Isolated Neutron Stars}},}\ }\href {\doibase 10.1086/171646}
  {\bibfield  {journal} {\bibinfo  {journal} {\apj}\ }\textbf {\bibinfo
  {volume} {395}},\ \bibinfo {pages} {250} (\bibinfo {year}
  {1992})}\BibitemShut {NoStop}%
\bibitem [{\citenamefont {{Gourgouliatos}}\ and\ \citenamefont
  {{Cumming}}(2014)}]{crust}%
  \BibitemOpen
  \bibfield  {author} {\bibinfo {author} {\bibfnamefont {K.~N.}\ \bibnamefont
  {{Gourgouliatos}}}\ and\ \bibinfo {author} {\bibfnamefont {A.}~\bibnamefont
  {{Cumming}}},\ }\bibfield  {title} {\enquote {\bibinfo {title} {{Hall effect
  in neutron star crusts: evolution, endpoint and dependence on initial
  conditions}},}\ }\href {\doibase 10.1093/mnras/stt2300} {\bibfield  {journal}
  {\bibinfo  {journal} {\mnras}\ }\textbf {\bibinfo {volume} {438}},\ \bibinfo
  {pages} {1618--1629} (\bibinfo {year} {2014})},\ \Eprint
  {http://arxiv.org/abs/1311.7004} {arXiv:1311.7004 [astro-ph.SR]} \BibitemShut
  {NoStop}%
\bibitem [{\citenamefont {{Castillo}}\ \emph {et~al.}(2017)\citenamefont
  {{Castillo}}, \citenamefont {{Reisenegger}},\ and\ \citenamefont
  {{Valdivia}}}]{core}%
  \BibitemOpen
  \bibfield  {author} {\bibinfo {author} {\bibfnamefont {F.}~\bibnamefont
  {{Castillo}}}, \bibinfo {author} {\bibfnamefont {A.}~\bibnamefont
  {{Reisenegger}}}, \ and\ \bibinfo {author} {\bibfnamefont {J.~A.}\
  \bibnamefont {{Valdivia}}},\ }\bibfield  {title} {\enquote {\bibinfo {title}
  {{Magnetic field evolution and equilibrium configurations in neutron star
  cores: the effect of ambipolar diffusion}},}\ }\href {\doibase
  10.1093/mnras/stx1604} {\bibfield  {journal} {\bibinfo  {journal} {\mnras}\
  }\textbf {\bibinfo {volume} {471}},\ \bibinfo {pages} {507--522} (\bibinfo
  {year} {2017})},\ \Eprint {http://arxiv.org/abs/1705.10020} {arXiv:1705.10020
  [astro-ph.HE]} \BibitemShut {NoStop}%
\bibitem [{\citenamefont {{Bransgrove}}\ \emph {et~al.}(2018)\citenamefont
  {{Bransgrove}}, \citenamefont {{Levin}},\ and\ \citenamefont
  {{Beloborodov}}}]{combined}%
  \BibitemOpen
  \bibfield  {author} {\bibinfo {author} {\bibfnamefont {Ashley}\ \bibnamefont
  {{Bransgrove}}}, \bibinfo {author} {\bibfnamefont {Yuri}\ \bibnamefont
  {{Levin}}}, \ and\ \bibinfo {author} {\bibfnamefont {Andrei}\ \bibnamefont
  {{Beloborodov}}},\ }\bibfield  {title} {\enquote {\bibinfo {title} {{Magnetic
  field evolution of neutron stars - I. Basic formalism, numerical techniques
  and first results}},}\ }\href {\doibase 10.1093/mnras/stx2508} {\bibfield
  {journal} {\bibinfo  {journal} {\mnras}\ }\textbf {\bibinfo {volume} {473}},\
  \bibinfo {pages} {2771--2790} (\bibinfo {year} {2018})},\ \Eprint
  {http://arxiv.org/abs/1709.09167} {arXiv:1709.09167 [astro-ph.HE]}
  \BibitemShut {NoStop}%
\bibitem [{\citenamefont {Thompson}\ and\ \citenamefont
  {Duncan}(1993)}]{thompson1993neutron}%
  \BibitemOpen
  \bibfield  {author} {\bibinfo {author} {\bibfnamefont {Christopher}\
  \bibnamefont {Thompson}}\ and\ \bibinfo {author} {\bibfnamefont {Robert~C}\
  \bibnamefont {Duncan}},\ }\bibfield  {title} {\enquote {\bibinfo {title}
  {Neutron star dynamos and the origins of pulsar magnetism},}\ }\href@noop {}
  {\bibfield  {journal} {\bibinfo  {journal} {The Astrophysical Journal}\
  }\textbf {\bibinfo {volume} {408}},\ \bibinfo {pages} {194--217} (\bibinfo
  {year} {1993})}\BibitemShut {NoStop}%
\bibitem [{\citenamefont {Thompson}\ and\ \citenamefont
  {Duncan}(1995)}]{thompson1995soft}%
  \BibitemOpen
  \bibfield  {author} {\bibinfo {author} {\bibfnamefont {Christopher}\
  \bibnamefont {Thompson}}\ and\ \bibinfo {author} {\bibfnamefont {Robert~C}\
  \bibnamefont {Duncan}},\ }\bibfield  {title} {\enquote {\bibinfo {title} {The
  soft gamma repeaters as very strongly magnetized neutron stars-i. radiative
  mechanism for outbursts},}\ }\href@noop {} {\bibfield  {journal} {\bibinfo
  {journal} {Monthly Notices of the Royal Astronomical Society}\ }\textbf
  {\bibinfo {volume} {275}},\ \bibinfo {pages} {255--300} (\bibinfo {year}
  {1995})}\BibitemShut {NoStop}%
\bibitem [{\citenamefont {Urpin}\ and\ \citenamefont
  {Shalybkov}(1999)}]{urpin1999magnetohydrodynamic}%
  \BibitemOpen
  \bibfield  {author} {\bibinfo {author} {\bibfnamefont {V}~\bibnamefont
  {Urpin}}\ and\ \bibinfo {author} {\bibfnamefont {D}~\bibnamefont
  {Shalybkov}},\ }\bibfield  {title} {\enquote {\bibinfo {title}
  {Magnetohydrodynamic processes in strongly magnetized young neutron stars},}\
  }\href@noop {} {\bibfield  {journal} {\bibinfo  {journal} {Monthly Notices of
  the Royal Astronomical Society}\ }\textbf {\bibinfo {volume} {304}},\
  \bibinfo {pages} {451--456} (\bibinfo {year} {1999})}\BibitemShut {NoStop}%
\bibitem [{\citenamefont {Cumming}\ \emph {et~al.}(2004)\citenamefont
  {Cumming}, \citenamefont {Arras},\ and\ \citenamefont
  {Zweibel}}]{Cumming:2004mf}%
  \BibitemOpen
  \bibfield  {author} {\bibinfo {author} {\bibfnamefont {Andrew}\ \bibnamefont
  {Cumming}}, \bibinfo {author} {\bibfnamefont {Phil}\ \bibnamefont {Arras}}, \
  and\ \bibinfo {author} {\bibfnamefont {Ellen~G.}\ \bibnamefont {Zweibel}},\
  }\bibfield  {title} {\enquote {\bibinfo {title} {{Magnetic field evolution in
  neutron star crusts due to the Hall effect and Ohmic decay}},}\ }\href
  {\doibase 10.1086/421324} {\bibfield  {journal} {\bibinfo  {journal}
  {Astrophys. J.}\ }\textbf {\bibinfo {volume} {609}},\ \bibinfo {pages}
  {999--1017} (\bibinfo {year} {2004})},\ \Eprint
  {http://arxiv.org/abs/astro-ph/0402392} {arXiv:astro-ph/0402392} \BibitemShut
  {NoStop}%
\bibitem [{\citenamefont {Cho}\ and\ \citenamefont
  {Lazarian}(2004)}]{cho2004anisotropy}%
  \BibitemOpen
  \bibfield  {author} {\bibinfo {author} {\bibfnamefont {Jungyeon}\
  \bibnamefont {Cho}}\ and\ \bibinfo {author} {\bibfnamefont {A}~\bibnamefont
  {Lazarian}},\ }\bibfield  {title} {\enquote {\bibinfo {title} {The anisotropy
  of electron magnetohydrodynamic turbulence},}\ }\href@noop {} {\bibfield
  {journal} {\bibinfo  {journal} {The Astrophysical Journal}\ }\textbf
  {\bibinfo {volume} {615}},\ \bibinfo {pages} {L41} (\bibinfo {year}
  {2004})}\BibitemShut {NoStop}%
\bibitem [{\citenamefont {Reisenegger}\ \emph {et~al.}(2005)\citenamefont
  {Reisenegger}, \citenamefont {Prieto}, \citenamefont {Benguria},
  \citenamefont {Lai},\ and\ \citenamefont {Araya}}]{Reisenegger:2005nb}%
  \BibitemOpen
  \bibfield  {author} {\bibinfo {author} {\bibfnamefont {Andreas}\ \bibnamefont
  {Reisenegger}}, \bibinfo {author} {\bibfnamefont {Joaquin}\ \bibnamefont
  {Prieto}}, \bibinfo {author} {\bibfnamefont {Rafael}\ \bibnamefont
  {Benguria}}, \bibinfo {author} {\bibfnamefont {Dong}\ \bibnamefont {Lai}}, \
  and\ \bibinfo {author} {\bibfnamefont {Pablo}\ \bibnamefont {Araya}},\
  }\bibfield  {title} {\enquote {\bibinfo {title} {{Magnetic fields in neutron
  stars: A Theoretical perspective}},}\ }\href {\doibase 10.1063/1.2077190}
  {\bibfield  {journal} {\bibinfo  {journal} {AIP Conf. Proc.}\ }\textbf
  {\bibinfo {volume} {784}},\ \bibinfo {pages} {263--273} (\bibinfo {year}
  {2005})},\ \Eprint {http://arxiv.org/abs/astro-ph/0503047}
  {arXiv:astro-ph/0503047} \BibitemShut {NoStop}%
\bibitem [{\citenamefont {Mereghetti}\ \emph {et~al.}(2015)\citenamefont
  {Mereghetti}, \citenamefont {Pons},\ and\ \citenamefont
  {Melatos}}]{Mereghetti:2015asa}%
  \BibitemOpen
  \bibfield  {author} {\bibinfo {author} {\bibfnamefont {Sandro}\ \bibnamefont
  {Mereghetti}}, \bibinfo {author} {\bibfnamefont {Jose'}\ \bibnamefont
  {Pons}}, \ and\ \bibinfo {author} {\bibfnamefont {Andrew}\ \bibnamefont
  {Melatos}},\ }\bibfield  {title} {\enquote {\bibinfo {title} {{Magnetars:
  Properties, Origin and Evolution}},}\ }\href {\doibase
  10.1007/s11214-015-0146-y} {\bibfield  {journal} {\bibinfo  {journal} {Space
  Sci. Rev.}\ }\textbf {\bibinfo {volume} {191}},\ \bibinfo {pages} {315--338}
  (\bibinfo {year} {2015})},\ \Eprint {http://arxiv.org/abs/1503.06313}
  {arXiv:1503.06313 [astro-ph.HE]} \BibitemShut {NoStop}%
\bibitem [{\citenamefont {Passamonti}\ \emph {et~al.}(2017)\citenamefont
  {Passamonti}, \citenamefont {Akg\"un}, \citenamefont {Pons},\ and\
  \citenamefont {Miralles}}]{Passamonti:2016nmf}%
  \BibitemOpen
  \bibfield  {author} {\bibinfo {author} {\bibfnamefont {Andrea}\ \bibnamefont
  {Passamonti}}, \bibinfo {author} {\bibfnamefont {Taner}\ \bibnamefont
  {Akg\"un}}, \bibinfo {author} {\bibfnamefont {Jos\'e~A.}\ \bibnamefont
  {Pons}}, \ and\ \bibinfo {author} {\bibfnamefont {Juan~A.}\ \bibnamefont
  {Miralles}},\ }\bibfield  {title} {\enquote {\bibinfo {title} {{The relevance
  of ambipolar diffusion for neutron star evolution}},}\ }\href {\doibase
  10.1093/mnras/stw2936} {\bibfield  {journal} {\bibinfo  {journal} {Mon. Not.
  Roy. Astron. Soc.}\ }\textbf {\bibinfo {volume} {465}},\ \bibinfo {pages}
  {3416--3428} (\bibinfo {year} {2017})},\ \Eprint
  {http://arxiv.org/abs/1608.00001} {arXiv:1608.00001 [astro-ph.HE]}
  \BibitemShut {NoStop}%
\bibitem [{\citenamefont {Pons}\ and\ \citenamefont
  {Vigan\`o}(2019)}]{Pons_2019}%
  \BibitemOpen
  \bibfield  {author} {\bibinfo {author} {\bibfnamefont {Jos\'e~A.}\
  \bibnamefont {Pons}}\ and\ \bibinfo {author} {\bibfnamefont {Daniele}\
  \bibnamefont {Vigan\`o}},\ }\bibfield  {title} {\enquote {\bibinfo {title}
  {Magnetic, thermal and rotational evolution of isolated neutron stars},}\
  }\href {\doibase 10.1007/s41115-019-0006-7} {\bibfield  {journal} {\bibinfo
  {journal} {Living Reviews in Computational Astrophysics}\ }\textbf {\bibinfo
  {volume} {5}} (\bibinfo {year} {2019}),\
  10.1007/s41115-019-0006-7}\BibitemShut {NoStop}%
\bibitem [{\citenamefont {Brandenburg}\ and\ \citenamefont
  {Zweibel}(1994)}]{Brandenburg1994TheFO}%
  \BibitemOpen
  \bibfield  {author} {\bibinfo {author} {\bibfnamefont {Axel}\ \bibnamefont
  {Brandenburg}}\ and\ \bibinfo {author} {\bibfnamefont {Ellen~G.}\
  \bibnamefont {Zweibel}},\ }\bibfield  {title} {\enquote {\bibinfo {title}
  {The formation of sharp structures by ambipolar diffusion},}\ }\href@noop {}
  {\bibfield  {journal} {\bibinfo  {journal} {The Astrophysical Journal}\
  }\textbf {\bibinfo {volume} {427}} (\bibinfo {year} {1994})}\BibitemShut
  {NoStop}%
\bibitem [{\citenamefont {Brandenburg}\ and\ \citenamefont
  {Subramanian}(2005)}]{brandenburg2005astrophysical}%
  \BibitemOpen
  \bibfield  {author} {\bibinfo {author} {\bibfnamefont {Axel}\ \bibnamefont
  {Brandenburg}}\ and\ \bibinfo {author} {\bibfnamefont {Kandaswamy}\
  \bibnamefont {Subramanian}},\ }\bibfield  {title} {\enquote {\bibinfo {title}
  {Astrophysical magnetic fields and nonlinear dynamo theory},}\ }\href@noop {}
  {\bibfield  {journal} {\bibinfo  {journal} {Physics Reports}\ }\textbf
  {\bibinfo {volume} {417}},\ \bibinfo {pages} {1--209} (\bibinfo {year}
  {2005})}\BibitemShut {NoStop}%
\bibitem [{\citenamefont {Balbus}(2009)}]{balbus2009magnetohydrodynamics}%
  \BibitemOpen
  \bibfield  {author} {\bibinfo {author} {\bibfnamefont {Steven~A.}\
  \bibnamefont {Balbus}},\ }\href@noop {} {\enquote {\bibinfo {title}
  {Magnetohydrodynamics of protostellar disks},}\ } (\bibinfo {year} {2009}),\
  \Eprint {http://arxiv.org/abs/0906.0854} {arXiv:0906.0854 [astro-ph.SR]}
  \BibitemShut {NoStop}%
\bibitem [{\citenamefont {Glampedakis}\ \emph {et~al.}(2010)\citenamefont
  {Glampedakis}, \citenamefont {Andersson},\ and\ \citenamefont
  {Samuelsson}}]{10.1111/j.1365-2966.2010.17484.x}%
  \BibitemOpen
  \bibfield  {author} {\bibinfo {author} {\bibfnamefont {Kostas}\ \bibnamefont
  {Glampedakis}}, \bibinfo {author} {\bibfnamefont {Nils}\ \bibnamefont
  {Andersson}}, \ and\ \bibinfo {author} {\bibfnamefont {Lars}\ \bibnamefont
  {Samuelsson}},\ }\bibfield  {title} {\enquote {\bibinfo {title}
  {{Magnetohydrodynamics of superfluid and superconducting neutron star
  cores}},}\ }\href {\doibase 10.1111/j.1365-2966.2010.17484.x} {\bibfield
  {journal} {\bibinfo  {journal} {Monthly Notices of the Royal Astronomical
  Society}\ }\textbf {\bibinfo {volume} {410}},\ \bibinfo {pages} {805--829}
  (\bibinfo {year} {2010})}\BibitemShut {NoStop}%
\bibitem [{\citenamefont {{Dommes}}\ and\ \citenamefont
  {{Gusakov}}(2021)}]{dommes}%
  \BibitemOpen
  \bibfield  {author} {\bibinfo {author} {\bibfnamefont {V.~A.}\ \bibnamefont
  {{Dommes}}}\ and\ \bibinfo {author} {\bibfnamefont {M.~E.}\ \bibnamefont
  {{Gusakov}}},\ }\bibfield  {title} {\enquote {\bibinfo {title} {{Dissipative
  superfluid relativistic magnetohydrodynamics of a multicomponent fluid: The
  combined effect of particle diffusion and vortices}},}\ }\href {\doibase
  10.1103/PhysRevD.104.123008} {\bibfield  {journal} {\bibinfo  {journal}
  {\prd}\ }\textbf {\bibinfo {volume} {104}},\ \bibinfo {eid} {123008}
  (\bibinfo {year} {2021})},\ \Eprint {http://arxiv.org/abs/2111.00999}
  {arXiv:2111.00999 [astro-ph.HE]} \BibitemShut {NoStop}%
\bibitem [{\citenamefont {{Vardhan}}\ \emph {et~al.}(2024)\citenamefont
  {{Vardhan}}, \citenamefont {{Grozdanov}}, \citenamefont {{Leutheusser}},\
  and\ \citenamefont {{Liu}}}]{long}%
  \BibitemOpen
  \bibfield  {author} {\bibinfo {author} {\bibfnamefont {Shreya}\ \bibnamefont
  {{Vardhan}}}, \bibinfo {author} {\bibfnamefont {Sa{\v{s}}o}\ \bibnamefont
  {{Grozdanov}}}, \bibinfo {author} {\bibfnamefont {Samuel}\ \bibnamefont
  {{Leutheusser}}}, \ and\ \bibinfo {author} {\bibfnamefont {Hong}\
  \bibnamefont {{Liu}}},\ }\bibfield  {title} {\enquote {\bibinfo {title}
  {{Effective field theories of dissipative fluids with one-form
  symmetries}},}\ }\href {\doibase 10.48550/arXiv.2408.12868} {\bibfield
  {journal} {\bibinfo  {journal} {arXiv e-prints}\ ,\ \bibinfo {pages}
  {arXiv:2408.12868}} (\bibinfo {year} {2024})},\ \Eprint
  {http://arxiv.org/abs/2408.12868} {arXiv:2408.12868 [hep-th]} \BibitemShut
  {NoStop}%
\bibitem [{\citenamefont {Crossley}\ \emph {et~al.}(2017)\citenamefont
  {Crossley}, \citenamefont {Glorioso},\ and\ \citenamefont
  {Liu}}]{Crossley:2015evo}%
  \BibitemOpen
  \bibfield  {author} {\bibinfo {author} {\bibfnamefont {Michael}\ \bibnamefont
  {Crossley}}, \bibinfo {author} {\bibfnamefont {Paolo}\ \bibnamefont
  {Glorioso}}, \ and\ \bibinfo {author} {\bibfnamefont {Hong}\ \bibnamefont
  {Liu}},\ }\bibfield  {title} {\enquote {\bibinfo {title} {{Effective field
  theory of dissipative fluids}},}\ }\href {\doibase 10.1007/JHEP09(2017)095}
  {\bibfield  {journal} {\bibinfo  {journal} {JHEP}\ }\textbf {\bibinfo
  {volume} {09}},\ \bibinfo {pages} {095} (\bibinfo {year} {2017})},\ \Eprint
  {http://arxiv.org/abs/1511.03646} {arXiv:1511.03646 [hep-th]} \BibitemShut
  {NoStop}%
\bibitem [{\citenamefont {Glorioso}\ \emph {et~al.}(2017)\citenamefont
  {Glorioso}, \citenamefont {Crossley},\ and\ \citenamefont
  {Liu}}]{Glorioso:2017fpd}%
  \BibitemOpen
  \bibfield  {author} {\bibinfo {author} {\bibfnamefont {Paolo}\ \bibnamefont
  {Glorioso}}, \bibinfo {author} {\bibfnamefont {Michael}\ \bibnamefont
  {Crossley}}, \ and\ \bibinfo {author} {\bibfnamefont {Hong}\ \bibnamefont
  {Liu}},\ }\bibfield  {title} {\enquote {\bibinfo {title} {{Effective field
  theory of dissipative fluids (II): classical limit, dynamical KMS symmetry
  and entropy current}},}\ }\href {\doibase 10.1007/JHEP09(2017)096} {\bibfield
   {journal} {\bibinfo  {journal} {JHEP}\ }\textbf {\bibinfo {volume} {09}},\
  \bibinfo {pages} {096} (\bibinfo {year} {2017})},\ \Eprint
  {http://arxiv.org/abs/1701.07817} {arXiv:1701.07817 [hep-th]} \BibitemShut
  {NoStop}%
\bibitem [{\citenamefont {Glorioso}\ and\ \citenamefont
  {Liu}(2018)}]{Glorioso:2018wxw}%
  \BibitemOpen
  \bibfield  {author} {\bibinfo {author} {\bibfnamefont {Paolo}\ \bibnamefont
  {Glorioso}}\ and\ \bibinfo {author} {\bibfnamefont {Hong}\ \bibnamefont
  {Liu}},\ }\bibfield  {title} {\enquote {\bibinfo {title} {{Lectures on
  non-equilibrium effective field theories and fluctuating hydrodynamics}},}\
  }\href@noop {} {\  (\bibinfo {year} {2018})},\ \Eprint
  {http://arxiv.org/abs/1805.09331} {arXiv:1805.09331 [hep-th]} \BibitemShut
  {NoStop}%
\bibitem [{\citenamefont {Grozdanov}\ and\ \citenamefont
  {Polonyi}(2015)}]{Grozdanov:2013dba}%
  \BibitemOpen
  \bibfield  {author} {\bibinfo {author} {\bibfnamefont {Sa\v{s}o}\
  \bibnamefont {Grozdanov}}\ and\ \bibinfo {author} {\bibfnamefont {Janos}\
  \bibnamefont {Polonyi}},\ }\bibfield  {title} {\enquote {\bibinfo {title}
  {{Viscosity and dissipative hydrodynamics from effective field theory}},}\
  }\href {\doibase 10.1103/PhysRevD.91.105031} {\bibfield  {journal} {\bibinfo
  {journal} {Phys. Rev.}\ }\textbf {\bibinfo {volume} {D91}},\ \bibinfo {pages}
  {105031} (\bibinfo {year} {2015})},\ \Eprint {http://arxiv.org/abs/1305.3670}
  {arXiv:1305.3670 [hep-th]} \BibitemShut {NoStop}%
\bibitem [{\citenamefont {Haehl}\ \emph {et~al.}(2016)\citenamefont {Haehl},
  \citenamefont {Loganayagam},\ and\ \citenamefont
  {Rangamani}}]{Haehl:2015uoc}%
  \BibitemOpen
  \bibfield  {author} {\bibinfo {author} {\bibfnamefont {Felix~M.}\
  \bibnamefont {Haehl}}, \bibinfo {author} {\bibfnamefont {R.}~\bibnamefont
  {Loganayagam}}, \ and\ \bibinfo {author} {\bibfnamefont {M.}~\bibnamefont
  {Rangamani}},\ }\bibfield  {title} {\enquote {\bibinfo {title} {{Topological
  sigma models \& dissipative hydrodynamics}},}\ }\href {\doibase
  10.1007/JHEP04(2016)039} {\bibfield  {journal} {\bibinfo  {journal} {JHEP}\
  }\textbf {\bibinfo {volume} {04}},\ \bibinfo {pages} {039} (\bibinfo {year}
  {2016})},\ \Eprint {http://arxiv.org/abs/1511.07809} {arXiv:1511.07809
  [hep-th]} \BibitemShut {NoStop}%
\bibitem [{\citenamefont {Jensen}\ \emph {et~al.}(2018)\citenamefont {Jensen},
  \citenamefont {Pinzani-Fokeeva},\ and\ \citenamefont
  {Yarom}}]{Jensen:2017kzi}%
  \BibitemOpen
  \bibfield  {author} {\bibinfo {author} {\bibfnamefont {Kristan}\ \bibnamefont
  {Jensen}}, \bibinfo {author} {\bibfnamefont {Natalia}\ \bibnamefont
  {Pinzani-Fokeeva}}, \ and\ \bibinfo {author} {\bibfnamefont {Amos}\
  \bibnamefont {Yarom}},\ }\bibfield  {title} {\enquote {\bibinfo {title}
  {{Dissipative hydrodynamics in superspace}},}\ }\href {\doibase
  10.1007/JHEP09(2018)127} {\bibfield  {journal} {\bibinfo  {journal} {JHEP}\
  }\textbf {\bibinfo {volume} {09}},\ \bibinfo {pages} {127} (\bibinfo {year}
  {2018})},\ \Eprint {http://arxiv.org/abs/1701.07436} {arXiv:1701.07436
  [hep-th]} \BibitemShut {NoStop}%
\bibitem [{\citenamefont {Grozdanov}\ \emph {et~al.}(2017)\citenamefont
  {Grozdanov}, \citenamefont {Hofman},\ and\ \citenamefont
  {Iqbal}}]{Grozdanov:2016tdf}%
  \BibitemOpen
  \bibfield  {author} {\bibinfo {author} {\bibfnamefont {Sa\v{s}o}\
  \bibnamefont {Grozdanov}}, \bibinfo {author} {\bibfnamefont {Diego~M.}\
  \bibnamefont {Hofman}}, \ and\ \bibinfo {author} {\bibfnamefont {Nabil}\
  \bibnamefont {Iqbal}},\ }\bibfield  {title} {\enquote {\bibinfo {title}
  {{Generalized global symmetries and dissipative magnetohydrodynamics}},}\
  }\href {\doibase 10.1103/PhysRevD.95.096003} {\bibfield  {journal} {\bibinfo
  {journal} {Phys. Rev.}\ }\textbf {\bibinfo {volume} {D95}},\ \bibinfo {pages}
  {096003} (\bibinfo {year} {2017})},\ \Eprint
  {http://arxiv.org/abs/1610.07392} {arXiv:1610.07392 [hep-th]} \BibitemShut
  {NoStop}%
\bibitem [{\citenamefont {Gaiotto}\ \emph {et~al.}(2015)\citenamefont
  {Gaiotto}, \citenamefont {Kapustin}, \citenamefont {Seiberg},\ and\
  \citenamefont {Willett}}]{Gaiotto:2014kfa}%
  \BibitemOpen
  \bibfield  {author} {\bibinfo {author} {\bibfnamefont {Davide}\ \bibnamefont
  {Gaiotto}}, \bibinfo {author} {\bibfnamefont {Anton}\ \bibnamefont
  {Kapustin}}, \bibinfo {author} {\bibfnamefont {Nathan}\ \bibnamefont
  {Seiberg}}, \ and\ \bibinfo {author} {\bibfnamefont {Brian}\ \bibnamefont
  {Willett}},\ }\bibfield  {title} {\enquote {\bibinfo {title} {{Generalized
  Global Symmetries}},}\ }\href {\doibase 10.1007/JHEP02(2015)172} {\bibfield
  {journal} {\bibinfo  {journal} {JHEP}\ }\textbf {\bibinfo {volume} {02}},\
  \bibinfo {pages} {172} (\bibinfo {year} {2015})},\ \Eprint
  {http://arxiv.org/abs/1412.5148} {arXiv:1412.5148 [hep-th]} \BibitemShut
  {NoStop}%
\bibitem [{\citenamefont {Hernandez}\ and\ \citenamefont
  {Kovtun}(2017)}]{Hernandez:2017mch}%
  \BibitemOpen
  \bibfield  {author} {\bibinfo {author} {\bibfnamefont {Juan}\ \bibnamefont
  {Hernandez}}\ and\ \bibinfo {author} {\bibfnamefont {Pavel}\ \bibnamefont
  {Kovtun}},\ }\bibfield  {title} {\enquote {\bibinfo {title} {{Relativistic
  magnetohydrodynamics}},}\ }\href {\doibase 10.1007/JHEP05(2017)001}
  {\bibfield  {journal} {\bibinfo  {journal} {JHEP}\ }\textbf {\bibinfo
  {volume} {05}},\ \bibinfo {pages} {001} (\bibinfo {year} {2017})},\ \Eprint
  {http://arxiv.org/abs/1703.08757} {arXiv:1703.08757 [hep-th]} \BibitemShut
  {NoStop}%
\bibitem [{\citenamefont {Grozdanov}\ and\ \citenamefont
  {Poovuttikul}(2019)}]{Grozdanov:2017kyl}%
  \BibitemOpen
  \bibfield  {author} {\bibinfo {author} {\bibfnamefont {Sa\v{s}o}\
  \bibnamefont {Grozdanov}}\ and\ \bibinfo {author} {\bibfnamefont {Napat}\
  \bibnamefont {Poovuttikul}},\ }\bibfield  {title} {\enquote {\bibinfo {title}
  {{Generalised global symmetries in holography: magnetohydrodynamic waves in a
  strongly interacting plasma}},}\ }\href {\doibase 10.1007/JHEP04(2019)141}
  {\bibfield  {journal} {\bibinfo  {journal} {JHEP}\ }\textbf {\bibinfo
  {volume} {04}},\ \bibinfo {pages} {141} (\bibinfo {year} {2019})},\ \Eprint
  {http://arxiv.org/abs/1707.04182} {arXiv:1707.04182 [hep-th]} \BibitemShut
  {NoStop}%
\bibitem [{\citenamefont {Glorioso}\ and\ \citenamefont
  {Son}(2018)}]{Glorioso:2018kcp}%
  \BibitemOpen
  \bibfield  {author} {\bibinfo {author} {\bibfnamefont {Paolo}\ \bibnamefont
  {Glorioso}}\ and\ \bibinfo {author} {\bibfnamefont {Dam~Thanh}\ \bibnamefont
  {Son}},\ }\bibfield  {title} {\enquote {\bibinfo {title} {{Effective field
  theory of magnetohydrodynamics from generalized global symmetries}},}\
  }\href@noop {} {\  (\bibinfo {year} {2018})},\ \Eprint
  {http://arxiv.org/abs/1811.04879} {arXiv:1811.04879 [hep-th]} \BibitemShut
  {NoStop}%
\bibitem [{\citenamefont {Armas}\ and\ \citenamefont
  {Jain}(2020)}]{Armas:2018zbe}%
  \BibitemOpen
  \bibfield  {author} {\bibinfo {author} {\bibfnamefont {Jay}\ \bibnamefont
  {Armas}}\ and\ \bibinfo {author} {\bibfnamefont {Akash}\ \bibnamefont
  {Jain}},\ }\bibfield  {title} {\enquote {\bibinfo {title} {{One-form
  superfluids \& magnetohydrodynamics}},}\ }\href {\doibase
  10.1007/JHEP01(2020)041} {\bibfield  {journal} {\bibinfo  {journal} {JHEP}\
  }\textbf {\bibinfo {volume} {01}},\ \bibinfo {pages} {041} (\bibinfo {year}
  {2020})},\ \Eprint {http://arxiv.org/abs/1811.04913} {arXiv:1811.04913
  [hep-th]} \BibitemShut {NoStop}%
\bibitem [{\citenamefont {Gralla}\ and\ \citenamefont
  {Iqbal}(2019)}]{Gralla:2018kif}%
  \BibitemOpen
  \bibfield  {author} {\bibinfo {author} {\bibfnamefont {Samuel~E.}\
  \bibnamefont {Gralla}}\ and\ \bibinfo {author} {\bibfnamefont {Nabil}\
  \bibnamefont {Iqbal}},\ }\bibfield  {title} {\enquote {\bibinfo {title}
  {{Effective Field Theory of Force-Free Electrodynamics}},}\ }\href {\doibase
  10.1103/PhysRevD.99.105004} {\bibfield  {journal} {\bibinfo  {journal} {Phys.
  Rev. D}\ }\textbf {\bibinfo {volume} {99}},\ \bibinfo {pages} {105004}
  (\bibinfo {year} {2019})},\ \Eprint {http://arxiv.org/abs/1811.07438}
  {arXiv:1811.07438 [hep-th]} \BibitemShut {NoStop}%
\bibitem [{\citenamefont {Benenowski}\ and\ \citenamefont
  {Poovuttikul}(2019)}]{Benenowski:2019ule}%
  \BibitemOpen
  \bibfield  {author} {\bibinfo {author} {\bibfnamefont {Bartosz}\ \bibnamefont
  {Benenowski}}\ and\ \bibinfo {author} {\bibfnamefont {Napat}\ \bibnamefont
  {Poovuttikul}},\ }\bibfield  {title} {\enquote {\bibinfo {title}
  {{Classification of magnetohydrodynamic transport at strong magnetic
  field}},}\ }\href@noop {} {\  (\bibinfo {year} {2019})},\ \Eprint
  {http://arxiv.org/abs/1911.05554} {arXiv:1911.05554 [hep-th]} \BibitemShut
  {NoStop}%
\bibitem [{\citenamefont {Landry}(2021)}]{Landry:2021kko}%
  \BibitemOpen
  \bibfield  {author} {\bibinfo {author} {\bibfnamefont {Michael~J.}\
  \bibnamefont {Landry}},\ }\bibfield  {title} {\enquote {\bibinfo {title}
  {{Higher-form and (non-)St\"uckelberg symmetries in non-equilibrium
  systems}},}\ }\href@noop {} {\  (\bibinfo {year} {2021})},\ \Eprint
  {http://arxiv.org/abs/2101.02210} {arXiv:2101.02210 [hep-th]} \BibitemShut
  {NoStop}%
\bibitem [{\citenamefont {Armas}\ and\ \citenamefont
  {Camilloni}(2022)}]{Armas:2022wvb}%
  \BibitemOpen
  \bibfield  {author} {\bibinfo {author} {\bibfnamefont {Jay}\ \bibnamefont
  {Armas}}\ and\ \bibinfo {author} {\bibfnamefont {Filippo}\ \bibnamefont
  {Camilloni}},\ }\bibfield  {title} {\enquote {\bibinfo {title} {{A stable and
  causal model of magnetohydrodynamics}},}\ }\href@noop {} {\  (\bibinfo {year}
  {2022})},\ \Eprint {http://arxiv.org/abs/2201.06847} {arXiv:2201.06847
  [hep-th]} \BibitemShut {NoStop}%
\bibitem [{Note3()}]{Note3}%
  \BibitemOpen
  \bibinfo {note} {Note that $\partial _0 G_{ij}= H_{0ij}+ \partial _i
  G_{0j}-\partial _j G_{0i}$.}\BibitemShut {Stop}%
\bibitem [{\citenamefont {Landry}\ and\ \citenamefont
  {Liu}(2022)}]{Landry:2022nog}%
  \BibitemOpen
  \bibfield  {author} {\bibinfo {author} {\bibfnamefont {Michael~J.}\
  \bibnamefont {Landry}}\ and\ \bibinfo {author} {\bibfnamefont {Hong}\
  \bibnamefont {Liu}},\ }\bibfield  {title} {\enquote {\bibinfo {title} {{A
  systematic formulation of chiral anomalous magnetohydrodynamics}},}\
  }\href@noop {} {\  (\bibinfo {year} {2022})},\ \Eprint
  {http://arxiv.org/abs/2212.09757} {arXiv:2212.09757 [hep-ph]} \BibitemShut
  {NoStop}%
\bibitem [{\citenamefont {Landry}\ and\ \citenamefont {Liu}()}]{LandryLiu}%
  \BibitemOpen
  \bibfield  {author} {\bibinfo {author} {\bibfnamefont {Michael~J.}\
  \bibnamefont {Landry}}\ and\ \bibinfo {author} {\bibfnamefont {Hong}\
  \bibnamefont {Liu}},\ }\bibfield  {title} {\enquote {\bibinfo {title} {{to
  appear}},}\ }\href@noop {} {\ }\Eprint {http://arxiv.org/abs/to appear} {to
  appear} \BibitemShut {NoStop}%
\bibitem [{Note4()}]{Note4}%
  \BibitemOpen
  \bibinfo {note} {In this equation, ${\protect \bf H}$ refers to the external
  magnetic field, which does not appear explicitly in our formalism, and in
  particular should not be confused with the higher form field
  $H_{ijk}$.}\BibitemShut {Stop}%
\bibitem [{\citenamefont {Hofman}\ and\ \citenamefont
  {Iqbal}(2018)}]{Hofman:2017vwr}%
  \BibitemOpen
  \bibfield  {author} {\bibinfo {author} {\bibfnamefont {Diego~M.}\
  \bibnamefont {Hofman}}\ and\ \bibinfo {author} {\bibfnamefont {Nabil}\
  \bibnamefont {Iqbal}},\ }\bibfield  {title} {\enquote {\bibinfo {title}
  {{Generalized global symmetries and holography}},}\ }\href {\doibase
  10.21468/SciPostPhys.4.1.005} {\bibfield  {journal} {\bibinfo  {journal}
  {SciPost Phys.}\ }\textbf {\bibinfo {volume} {4}},\ \bibinfo {pages} {005}
  (\bibinfo {year} {2018})},\ \Eprint {http://arxiv.org/abs/1707.08577}
  {arXiv:1707.08577 [hep-th]} \BibitemShut {NoStop}%
\bibitem [{\citenamefont {{Vigan{\`o}}}\ \emph {et~al.}(2019)\citenamefont
  {{Vigan{\`o}}}, \citenamefont {{Mart{\'\i}nez-G{\'o}mez}}, \citenamefont
  {{Pons}}, \citenamefont {{Palenzuela}}, \citenamefont {{Carrasco}},
  \citenamefont {{Mi{\~n}ano}}, \citenamefont {{Arbona}}, \citenamefont
  {{Bona}},\ and\ \citenamefont {{Mass{\'o}}}}]{vigano2}%
  \BibitemOpen
  \bibfield  {author} {\bibinfo {author} {\bibfnamefont {Daniele}\ \bibnamefont
  {{Vigan{\`o}}}}, \bibinfo {author} {\bibfnamefont {David}\ \bibnamefont
  {{Mart{\'\i}nez-G{\'o}mez}}}, \bibinfo {author} {\bibfnamefont {Jos{\'e}~A.}\
  \bibnamefont {{Pons}}}, \bibinfo {author} {\bibfnamefont {Carlos}\
  \bibnamefont {{Palenzuela}}}, \bibinfo {author} {\bibfnamefont {Federico}\
  \bibnamefont {{Carrasco}}}, \bibinfo {author} {\bibfnamefont {Borja}\
  \bibnamefont {{Mi{\~n}ano}}}, \bibinfo {author} {\bibfnamefont {Antoni}\
  \bibnamefont {{Arbona}}}, \bibinfo {author} {\bibfnamefont {Carles}\
  \bibnamefont {{Bona}}}, \ and\ \bibinfo {author} {\bibfnamefont {Joan}\
  \bibnamefont {{Mass{\'o}}}},\ }\bibfield  {title} {\enquote {\bibinfo {title}
  {{A Simflowny-based high-performance 3D code for the generalized induction
  equation}},}\ }\href {\doibase 10.1016/j.cpc.2018.11.022} {\bibfield
  {journal} {\bibinfo  {journal} {Computer Physics Communications}\ }\textbf
  {\bibinfo {volume} {237}},\ \bibinfo {pages} {168--183} (\bibinfo {year}
  {2019})},\ \Eprint {http://arxiv.org/abs/1811.08198} {arXiv:1811.08198
  [astro-ph.IM]} \BibitemShut {NoStop}%
\bibitem [{Note5()}]{Note5}%
  \BibitemOpen
  \bibinfo {note} {Note that we need an initial configuration which involves
  both $k_z$ and $k_\perp $ to probe this difference}\BibitemShut {NoStop}%
\bibitem [{\citenamefont {{Burge}}\ \emph {et~al.}(2016)\citenamefont
  {{Burge}}, \citenamefont {{Van Loo}}, \citenamefont {{Falle}},\ and\
  \citenamefont {{Hartquist}}}]{filament}%
  \BibitemOpen
  \bibfield  {author} {\bibinfo {author} {\bibfnamefont {C.~A.}\ \bibnamefont
  {{Burge}}}, \bibinfo {author} {\bibfnamefont {S.}~\bibnamefont {{Van Loo}}},
  \bibinfo {author} {\bibfnamefont {S.~A.~E.~G.}\ \bibnamefont {{Falle}}}, \
  and\ \bibinfo {author} {\bibfnamefont {T.~W.}\ \bibnamefont {{Hartquist}}},\
  }\bibfield  {title} {\enquote {\bibinfo {title} {{Ambipolar diffusion
  regulated collapse of filaments threaded by perpendicular magnetic
  fields}},}\ }\href {\doibase 10.1051/0004-6361/201629039} {\bibfield
  {journal} {\bibinfo  {journal} {\aap}\ }\textbf {\bibinfo {volume} {596}},\
  \bibinfo {eid} {A28} (\bibinfo {year} {2016})},\ \Eprint
  {http://arxiv.org/abs/1609.06879} {arXiv:1609.06879 [astro-ph.GA]}
  \BibitemShut {NoStop}%
\bibitem [{\citenamefont {{Ofengeim}}\ and\ \citenamefont
  {{Gusakov}}(2018)}]{ofengeim}%
  \BibitemOpen
  \bibfield  {author} {\bibinfo {author} {\bibfnamefont {D.~D.}\ \bibnamefont
  {{Ofengeim}}}\ and\ \bibinfo {author} {\bibfnamefont {M.~E.}\ \bibnamefont
  {{Gusakov}}},\ }\bibfield  {title} {\enquote {\bibinfo {title} {{Fast
  magnetic field evolution in neutron stars: The key role of magnetically
  induced fluid motions in the core}},}\ }\href {\doibase
  10.1103/PhysRevD.98.043007} {\bibfield  {journal} {\bibinfo  {journal}
  {\prd}\ }\textbf {\bibinfo {volume} {98}},\ \bibinfo {eid} {043007} (\bibinfo
  {year} {2018})},\ \Eprint {http://arxiv.org/abs/1805.03956} {arXiv:1805.03956
  [astro-ph.HE]} \BibitemShut {NoStop}%
\bibitem [{\citenamefont {{Castillo}}\ \emph {et~al.}(2020)\citenamefont
  {{Castillo}}, \citenamefont {{Reisenegger}},\ and\ \citenamefont
  {{Valdivia}}}]{castillovelocity}%
  \BibitemOpen
  \bibfield  {author} {\bibinfo {author} {\bibfnamefont {F.}~\bibnamefont
  {{Castillo}}}, \bibinfo {author} {\bibfnamefont {A.}~\bibnamefont
  {{Reisenegger}}}, \ and\ \bibinfo {author} {\bibfnamefont {J.~A.}\
  \bibnamefont {{Valdivia}}},\ }\bibfield  {title} {\enquote {\bibinfo {title}
  {{Two-fluid simulations of the magnetic field evolution in neutron star cores
  in the weak-coupling regime}},}\ }\href {\doibase 10.1093/mnras/staa2543}
  {\bibfield  {journal} {\bibinfo  {journal} {\mnras}\ }\textbf {\bibinfo
  {volume} {498}},\ \bibinfo {pages} {3000--3012} (\bibinfo {year} {2020})},\
  \Eprint {http://arxiv.org/abs/2006.13186} {arXiv:2006.13186 [astro-ph.HE]}
  \BibitemShut {NoStop}%
\end{thebibliography}%

\end{document}